%% file: ConicalAdS_final.tex
\documentclass[a4paper,11pt]{article}
\usepackage{jheppub}

\usepackage[utf8]{inputenc}
\usepackage[english]{babel}
\usepackage{mathtools}
\usepackage{amsfonts}
\usepackage{amsthm}
\usepackage{physics} 
\usepackage{comment} 
\usepackage[dvipsnames]{xcolor}
\usepackage{float} 
\usepackage{tikz} 
\usetikzlibrary{shapes,arrows,positioning,automata,backgrounds,calc,er,patterns}
\usepackage[compat=1.0.0]{tikz-feynman}
\usepackage{graphicx}

\usepackage{bbm}
\newcommand{\one}{\mathbbm{1}} 

\renewcommand{\a}{\alpha}
\renewcommand{\b}{\beta}
\newcommand{\g}{\gamma}
\renewcommand{\d}{\delta}
\newcommand{\eps}{\varepsilon}
\renewcommand{\th}{\theta}
\renewcommand{\k}{\kappa}
\renewcommand{\l}{\lambda}
\newcommand{\s}{\sigma}
\renewcommand{\t}{\tau}
\newcommand{\w}{\omega}
\newcommand{\G}{\Gamma}

\newcommand{\N}{\mathbb{N}} 
\newcommand{\Z}{\mathbb{Z}} 
\newcommand{\R}{\mathbb{R}} 

\newcommand{\DeclareAutoPairedDelimiter}[3]{%
  \expandafter\DeclarePairedDelimiter\csname Auto\string#1\endcsname{#2}{#3}%
  \begingroup\edef\x{\endgroup
    \noexpand\DeclareRobustCommand{\noexpand#1}{%
      \expandafter\noexpand\csname Auto\string#1\endcsname*}}%
  \x}
\DeclareAutoPairedDelimiter{\p}{(}{)} 

\newcommand{\too}{\longrightarrow} 
\renewcommand{\bar}[1]{\mkern1mu\overline{\mkern-1mu#1\mkern-1mu}\mkern1mu}
\renewcommand{\tilde}[1]{\widetilde{#1}}
\renewcommand{\L}{\mathcal{L}} 

\title{\boldmath Aspects of Holography in Conical $\mathrm{AdS}_3$}

\author{David Berenstein, David Grabovsky, and Ziyi Li}
\affiliation{Department of Physics, University of California at Santa Barbara, California 93106, USA}

\emailAdd{dberens@physics.ucsb.edu}
\emailAdd{davidgrabovsky@physics.ucsb.edu}
\emailAdd{ziyili@ucsb.edu}

\abstract{
We study the Feynman propagator of free scalar fields in $\mathrm{AdS}_3$ with a conical defect. In the bulk, the defect is represented by a massive particle; in the dual CFT, it is a heavy operator that creates a highly excited state. We construct the propagator by solving the bulk equation of motion in the defect geometry, summing over the modes of the field, and passing to the boundary. The result agrees with a calculation based on the method of images in $\mathrm{AdS}_3/\mathbb{Z}_N$, where it is also a sum over geodesic lengths. On the boundary, the propagator becomes a semiclassical heavy-light four-point function. We interpret the field modes as double-twist primary states formed by excitations of the scalar on top of the defect, and we check that the correlator is crossing-symmetric by matching its singular behavior to that of the semiclassical Virasoro vacuum block. We also argue that long-range correlations in conical AdS are ``thermally'' suppressed as the defect becomes more massive by studying the critical behavior of a continuous phase transition in the correlator at the BTZ threshold. Finally, we apply our results to holographic entanglement entropy by exploiting an analogy between free scalars and replica twist fields.
}

\begin{document} 
\maketitle
\flushbottom

\section{Introduction and Overview}

Holographic dualities \cite{Maldacena:1997re,Gubser:1998bc,Witten:1998qj} have proven to be a useful tool for exploring quantum field theories at strong coupling, where certain aspects of their physics often become universal. For example, thermal states in the boundary field theory are dual to black holes in a semiclassical bulk theory of quantum gravity. In the bulk, the Hawking--Page transition between thermal AdS and a black hole spacetime \cite{Hawking:1982dh} is usually a first-order phase transition that can be identified in the boundary theory \cite{Witten:1998zw}. But in $2+1$ dimensions, one finds a continuous phase transition in the microcanonical ensemble, because there is no minimal temperature for a non-rotating BTZ black hole. Instead, at a critical threshold energy there is a ``massless'' BTZ geometry of vanishing horizon size. Above the transition we have a black hole---how does the physics behave below the BTZ threshold?

\paragraph{Gravity with defects.} Without knowing the details, we can ask to what extent purely gravitational solutions might give us a guide to that physics. From one perspective, there are essentially no such solutions, because pure gravity in $2+1$ dimensions has no local propagating gravitational degrees of freedom \cite{Deser:1983nh}. However, the theory can contain matter concentrated on scales much smaller than the $\mathrm{AdS}$ radius without producing a black hole. (For example, in string theories one can have D-branes or other heavy objects whose size is the string scale.) As long as the mass of the matter source is below the BTZ threshold, no horizon will form. When the scales are very different, we can take the source to be pointlike. Such solutions can be purely gravitational away from the matter: this is allowed in 3 dimensions, where gravitational fields are essentially localized at their sources. In higher-dimensional setups, such configurations would necessarily form black holes.

In this paper, we consider a massive particle at the center of $\mathrm{AdS}_3$, represented in the bulk by a conical defect spacetime. The defect is labeled by a parameter $N \geq 1$ that determines its strength: the case $N=1$ corresponds to pure $\mathrm{AdS}_3$; integer values of $N$ describe the quotients $\mathrm{AdS}_3/\Z_N$; and in the limit $N \too \infty$ the particle reaches the BTZ threshold and becomes a black hole \cite{Banados:1992wn}. The conical defect metric tells us how gravity responds to the particle; we can also ask how other light fields in the bulk would see it. In this simplified model, we can bury the microscopic details under the rug: the model's simplicity allows us to study the physics of probe fields explicitly and analytically. In the end, we will find that a conical defect shares some similarities with black holes without actually being one, and we will explain how the physics of conical defects interpolates between the behavior of empty $\mathrm{AdS}$ and the thermal properties of black holes.

\paragraph{Scalars in conical AdS.} To probe the defect, we study free scalars in conical $\mathrm{AdS}_3$. We construct the scalar boundary-to-boundary propagator from the bulk equations of motion and use holographic techniques to discuss its properties. We devote particular attention to its characterization by the method of images and in terms of geodesics in $\mathrm{AdS}_3/\Z_N$, its interpretation and singular behavior in the dual conformal field theory (CFT), and its critical behavior as the defect's mass approaches the black hole threshold. 

Field theory in $\mathrm{AdS}_3$ is well understood \cite{Witten:1998qj, Avis:1977yn, Balasubramanian:1998sn, Freedman:1998tz, Berenstein:1998ij}, and the literature on scalars in BTZ is similarly extensive. Moreover, the propagator in $\mathrm{AdS}_3/\Z_N$ has been given an elegant treatment by the method of images \cite{Arefeva:2014aoe}. But most conical geometries correspond to non-integer values of $N$ and cannot be handled by the method of images. For these spacetimes, we construct the bulk propagator in canonical quantization, by summing directly over the modes of the field. One of our goals is to fill in the gaps between the integers and understand the dependence of the propagator on the defect's mass.

\paragraph{CFT interpretation.} We also aim to say something useful about the interpretation of the conical defect in the dual field theory. Our first step is to regard both the scalar and the defect as states in the CFT via the operator-state correspondence. Since the defect stays at the center of $\mathrm{AdS}$, it is a primary state; its conformal dimension is given by its mass. Moreover, the defect's lack of angular momentum makes it a scalar operator.

In AdS/CFT, correlators on the boundary can be obtained by taking regularized limits of the corresponding bulk correlators \cite{Witten:1998qj, Banks:1998dd}. In our case, the bulk propagator becomes a ``heavy-heavy-light-light'' four-point function involving the defect and the scalar. This correlator encodes the excitations of the light field in the bulk, and can be used to describe additional primary states built on top of the conical defect. It also encodes the operator product expansion (OPE) coefficients relating these excitations to the conical defect and the light field. In this sense, we can use the conical $\mathrm{AdS}_3$ propagator to learn about certain heavy states in AdS and their OPE coefficients. This discussion takes place in the s channel; we also do the t-channel calculation, which is dominated by the Virasoro vacuum block. This block has been studied extensively \cite{Fitzpatrick:2014vua,Fitzpatrick:2015zha} and captures gravitational corrections to the pure $\mathrm{AdS}_3$ propagator. Consistency between the two channels is important in understanding the normalization of the propagator.

\paragraph{Effective field theory.} One premise of our work is that the conical defect is so heavy that any other fields in its presence can be treated as free fluctuations that do not backreact on the geometry. (The defect mass is of order the central charge $c$, which we take to be large, and we take any other excitations on top of the defect to have energies of order one.) This premise justifies the study of non-interacting fields and allows us to compute the linear response of the defect to the small perturbations introduced by the scalar.

Since the defect makes fields in its presence approximately free, many complicated correlators should factorize and simplify dramatically. For instance, the entanglement entropy of a boundary interval is computed in the CFT from a certain correlator of replica twist fields \cite{Calabrese:2009qy}. By modeling the twist fields by quasi-free scalars, we can reproduce known results for the entanglement entropy in several cases. However, a subtle order-of-limits ambiguity prevents our proposal from computing the entropy more accurately.

\paragraph{Thermal behavior.} The qualitative behavior of field theories in black hole backgrounds, which are thermal systems, differs markedly from their behavior below the BTZ threshold. It is natural to ask whether conical defects approach thermality as $N$ increases, and whether the scalar can detect a phase transition between the conical and BTZ regimes. This paper answers both questions in the affirmative: we argue that long-range correlations of the scalar are screened by the defect, decaying monotonically in $N$; and we quantify the critical behavior of the correlator as one approaches the BTZ threshold from below.

\paragraph{Related work.} Recent work in several directions has demonstrated the importance of conical defects. The quotients $\mathrm{AdS}_3/\Z_N$ appear naturally in the replica trick for gravitational entropy \cite{Lewkowycz:2013nqa}; as contributions to the path integral of 3D gravity \cite{Benjamin:2020mfz}; as solutions of the D1-D5 system in string theory \cite{Martinec:2001cf,Balasubramanian:2005qu}, first identified by Lunin and Mathur \cite{Lunin:2001jy}; and in connection with an entropy-like quantity called entwinement \cite{Balasubramanian:2014sra}. Further work by Giusto, Russo, and others has also studied the more general D1-D5-P states \cite{Giusto:2004ip, Giusto:2012yz}. The heavy-light correlators in the geometries dual to these states are computed by studying the bulk wave equation \cite{Galliani:2016cai, Giusto:2020mup}, and also by methods developed in \cite{Kulaxizi:2018dxo, Karlsson:2019qfi}. Related work on the propagator in conical $\mathrm{AdS}_3$ has also been done by Arefeva and collaborators \cite{Arefeva:2014aoe, Ageev:2015qbz, Arefeva:2016nic, Arefeva:2016wek}. In several respects, we extend their analysis and improve on some of their findings. Whereas \cite{Arefeva:2016wek} focuses on scalars of integer dimension $\Delta \in \Z_+$, we obtain explicit results for any $\Delta \geq 1$ and make some remarks on the range $\Delta \in (0,1)$.
We also indicate how one should analytically continue results in $\mathrm{AdS}_3/\Z_N$ to arbitrary $N \in \R_+$.

For a complementary account of the CFT side of the story, see also \cite{Collier:2018exn}.

\paragraph{Outline.} We begin in \S\ref{sec:scalar-propagator} by solving the scalar's equation of motion. We impose suitable boundary conditions, normalize the field modes, canonically quantize the theory, and obtain the boundary Feynman propagator. In \S\ref{sec:properties}, we explore the singular behavior of the propagator, study it in $\mathrm{AdS}_3/\Z_N$ using the method of images, and interpret the result as a sum over geodesic lengths. In \S\ref{sec:CFT-interpretation}, we discuss our results in the language of the boundary CFT, and we compute the heavy-light four-point function in both the s- and t-channel OPEs. In \S\ref{sec:thermal-physics}, we argue that increasing the mass of the defect towards the BTZ threshold signals an approach to thermal physics. In particular, we show that the propagator experiences a continuous phase transition at the BTZ threshold. Finally, in \S\ref{sec:applications} we apply our results to holographic entanglement entropy. We argue that, in some regimes, one can extract the entropy of an interval in conical $\mathrm{AdS}_3$ directly from the scalar propagator. 
We close with a discussion of some interesting questions raised by our work in \S\ref{sec:conclusions}.

\section{\boldmath The Scalar Propagator in Conical $\mathrm{AdS}_3$}
\label{sec:scalar-propagator}

Consider a free scalar field in conical $\mathrm{AdS}_3$. We will begin by solving the equation of motion $(\Box - m^2)\phi(x) = 0$ to obtain field modes $\phi_i(x)$ that satisfy suitable boundary conditions. After orthonormalizing these modes with respect to the appropriate inner product, we will quantize the theory by imposing the standard equal-time commutation relations:
\begin{align}
\label{eqn:commutators}
\phi(x) = \sum_{i} \Big( a_i \phi_i(x) + a_i^{\dagger} \phi_i^*(x) \Big), \qquad [a_i,\, a^{\dagger}_j] = \d_{ij}, \quad [a_i,\, a_j] = [a_i^{\dagger},\, a_j^{\dagger}] = 0.
\end{align}
These relations serve to define the ground state $\ket{0}_N$ of the theory. They allow us to construct the bulk Wightman functions $G^{\pm}_{N}(x,x')$, which are defined by
\begin{align}
\label{eqn:wightman0}
G^+_{N}(x,x') = {}_{N}\!\mel{0}{\phi(x) \phi(x')}{0}_N = \sum_i \phi_i(x) \phi_i^*(x'), \qquad G^-_{N}(x,x') = G^+_{N}(x',x).
\end{align}
Finally, we will use the holographic extrapolate dictionary to send $x$ and $x'$ to the $\mathrm{AdS}$ boundary and assemble the boundary Feynman propagator.

\subsection{The Conical Defect Spacetime}
\label{sec:conical-defect}

In global coordinates $x = (t, r, \th)$, the conical $\mathrm{AdS}_3$ spacetime has the following metric:\footnote{Here and throughout, we set the AdS radius $\ell_{\mathrm{AdS}}$ to 1. We also work in units where $c = \hbar = 1$.}
\begin{align}
\label{eqn:metric}
\dd s^2 = 
-\p{r^2 + \frac{1}{N^2}} \dd t^2 + \p{r^2 + \frac{1}{N^2}}^{-1} \dd r^2 + r^2\, \dd\th^2, \qquad
N \in \R_+.
\end{align}
Here $t \in \R$ represents time, $r \in \R_+$ is a radial coordinate, and $\th \in [0, 2\pi)$ is the angular direction. The parameter $N$ measures the strength of the defect, with $N=1$ corresponding to pure $\mathrm{AdS}_3$. (Later on we will also use $\a = \frac{1}{N} \in (0,1]$ instead of $N$.) In the coordinates $(T,R,\phi)$ defined by $t = NT$, $r = \frac{R}{N}$, and $\th = N\phi$, the metric reduces to that of pure $\mathrm{AdS}_3$, except that $\phi \sim \phi + \frac{2\pi}{N}$ experiences an angular deficit of $\d\phi = 2\pi\big( 1 - \frac{1}{N} \big)$. The spacetime is viewed as a stack of cones, with a static particle at the tip of the cone. In the dual CFT, the particle is represented by a heavy primary operator. The positivity of its conformal dimension, required by unitarity, ensures that $N \geq 1$ and rules out conical excesses.

When $N$ is an integer, the metric (\ref{eqn:metric}) describes the quotient spacetime $\mathrm{AdS}_3/\Z_N$. As $N$ approaches $\infty$, the mass $M$ of the defect approaches the BTZ threshold $M_* = \frac{1}{8G}$: this is the massless BTZ geometry, a black hole with zero horizon size. If one takes $N = \frac{i}{r_+}$ to be imaginary, the metric becomes that of a BTZ black hole with a horizon at $r = r_+$.

\subsection{Solution of the Equation of Motion}
\label{sec:equations-of-motion}

In coordinates $(t,r,\th)$, the Klein--Gordon equation $(\Box - m^2)\phi = 0$ reads
\begin{align}
\label{eqn:eom}
\Bigg[\!-\Big(r^2 + \frac{1}{N^2}\Big)^{-1} \partial_t^2 + \frac{1}{r} \partial_r \p{r \Big(r^2 + \frac{1}{N^2}\Big) \partial_r} + \frac{1}{r^2} \partial_{\th}^2 - m^2 \Bigg] \phi(t,r,\th) = 0.
\end{align}
The Killing symmetries of the metric (\ref{eqn:metric}) make this equation separable, with plane-wave solutions in $t$ and $\th$ labeled by energy $\w$ and angular momentum $\ell$, respectively:
\begin{align}
\label{eqn:separable}
\phi_i(x) = \phi_{\w\ell}(t,r,\th) = e^{-i\w t} e^{i\ell \th} R_{\w\ell}(r), \qquad \w \in \R, \quad \ell \in \Z.
\end{align}
The angular momentum is quantized as an integer due to the periodicity of the $\th$ direction, since $\th \sim \th + 2\pi$, while the energy is---for now---unconstrained. After substituting the ansatz (\ref{eqn:separable}) into (\ref{eqn:eom}), we obtain a Schr\"odinger-like equation for the radial modes:
\begin{align}
\label{eqn:radial-equation}
\Bigg[ \frac{1}{r} \partial_r \p{r \Big(r^2 + \frac{1}{N^2}\Big) \partial_r} + \Big(r^2 + \frac{1}{N^2}\Big)^{-1} \w^2 - \frac{\ell^2}{r^2} - m^2 \Bigg] R_{\w\ell}(r) = 0.
\end{align}
The solutions to (\ref{eqn:radial-equation}) are hypergeometric functions. They are most naturally expressed in terms of $\Delta = 1 + \sqrt{1 + m^2}$, the conformal dimension of the CFT operator dual to $\phi$:
\begin{equation}
\begin{aligned}
\label{eqn:radial-solutions}
R_{\w\ell}(r) &= \frac{1}{N} \p{r^2 + \frac{1}{N^2}}^{N\w/2} \Bigg[ c^{+}_{\w\ell} r^{N\ell} \mathcal{F}^{+}_{\w\ell}(r) + c^{-}_{\w\ell} e^{-2\pi i N\ell} (N^2 r)^{-N\ell} \mathcal{F}^{-}_{\w\ell}(r) \Bigg], \\
\mathcal{F}^{\pm}_{\w\ell}(r) &= {}_2 F_1 \Big[ \frac{1}{2} \big( N\w \pm N\ell + 2 - \Delta\big),\, \frac{1}{2} \big( N\w \pm N\ell + \Delta\big);\, 1 \pm N\ell;\, -N^2 r^2 \Big].
\end{aligned}
\end{equation}

To fix the values of $c^{\pm}_{\w\ell}$, we must ensure that the modes are finite and regular at $r=0$, and that they are normalizable.\footnote{Strictly speaking, admissible solutions $\phi_{\w\ell}$ must have finite action. This is a stronger condition than the normalizability of the field profile \cite{Klebanov:1999tb}, but the two requirements are identical for individual modes.} For the first condition, we have $\mathcal{F}_{\w\ell}^{\pm}(0) = 1$, so near the origin $R_{\w\ell}(r) \sim c^{+}_{\w\ell} r^{N\ell} + c^{-}_{\w\ell} r^{-N\ell}$. When $\ell > 0$, the second term diverges as $r \too 0$, so we set $c_{\w\ell}^- = 0$ to avoid a singularity. Similarly, when $\ell < 0$, we take $c_{\w\ell}^+ = 0$.\footnote{The case $\ell = 0$ looks special, since both terms in $R_{\w 0} \sim c^+_{\w 0} + c^-_{\w 0}$ are regular at the origin. But the solutions to (\ref{eqn:radial-equation}) with $\ell = 0$, when handled separately, are still consistent with the result (\ref{eqn:mode-solutions}).} Thus we always keep the $r^{N|\ell|}$ term, which leaves us with the normalizable modes
\begin{equation}
\label{eqn:mode-solutions}
\begin{aligned}
\phi_{\w\ell}(t,r,\th) &= \mathcal{N}_{\w\ell}\, e^{i\w t} e^{i\ell\th} \p{r^2 + \frac{1}{N^2}}^{N\w/2} r^{N|\ell|} \mathcal{F}_{\w\ell}(r), \\
\mathcal{F}_{\w\ell}(r) &= {}_2 F_1 \Big[ \frac{1}{2} \big( N\w + N|\ell| + 2 - \Delta\big),\, \frac{1}{2} \big( N\w + N|\ell| + \Delta\big);\, 1 + N|\ell|;\, -N^2 r^2 \Big].
\end{aligned}
\end{equation}
Here $\mathcal{N}_{\w\ell} \sim \frac{1}{N} c_{\w\ell}^{\pm}$ is the normalization constant: our next goal is to find its value.

\subsection{Normalizability of the Modes}
\label{sec:normalizability}

The issue of normalization is more subtle. The modes (\ref{eqn:mode-solutions}) must be orthonormal in the Klein-Gordon inner product on the space of solutions to the equations of motion:
\begin{align}
\label{eqn:KG-product}
\big\langle \phi_i,\, \phi_j \big\rangle = -i \int_{\Sigma} \dd^2 x \sqrt{g_{\Sigma}(x)}\, \hat{n}^{\mu} \phi_i(x) \overset{\leftrightarrow}{\partial}_{\!\mu} \phi_j^*(x) = \d_{ij}, \quad \big\Vert \phi_{\w\ell} \big\Vert^2 = \big\langle \phi_{\w\ell},\, \phi_{\w\ell} \big\rangle = 1.
\end{align}
Here $\Sigma$ is any hypersurface of constant $t$, $g_{\Sigma}$ is the induced metric on $\Sigma$, and $\hat{n}^{\mu}$ is the future-directed unit normal to $\Sigma$. In particular, we have $\sqrt{g_{\Sigma}(x)} = r \p{r^2 + \frac{1}{N^2}}^{-1/2}$ and $\hat{n}^{\mu} = \frac{1}{\sqrt{-g_{tt}}} \p{\partial_t}^{\mu} = \p{r^2 + \frac{1}{N^2}}^{-1/2} \delta^{\mu t}$. Substituting these and (\ref{eqn:mode-solutions}) into (\ref{eqn:KG-product}), we find
\begin{align}
\label{eqn:mode-norm}
\big\Vert \phi_{\w\ell} \big\Vert^2 = 4\pi \w \int_0^{\infty} \dd r\, r \p{r^2 + \frac{1}{N^2}}^{-1} \big| R_{\w\ell}(r) \big|^2 = 1.
\end{align}

Before performing this integral to fix $\mathcal{N}_{\w\ell}$, we observe that at large $r$, the integrand behaves like $\frac{1}{r} \abs{R_{\w\ell}(r)}^2$. So in order for $\big\Vert \phi_{\w\ell} \big\Vert^2$ to be finite, $R_{\w\ell}(r)$ must vanish at infinity. However, an expansion of $R_{\w\ell}(r)$ in small $\frac{1}{r}$ shows that this is not always the case:
\begin{equation}
\begin{aligned}
\label{eqn:asymptotics}
R_{\w\ell}(r) \sim \mathcal{N}_{\w\ell} \Bigg[A^+_{\w\ell}(\Delta)\, r^{\Delta - 2} \p{1 + O\Big(\frac{1}{r^2}\Big)} + A^-_{\w\ell}(\Delta)\, r^{-\Delta} \p{1 + O\Big(\frac{1}{r^2}\Big)} \Bigg].
\end{aligned}
\end{equation}
The coefficients $A^{\pm}_{\w\ell}(\Delta)$ above can be expressed in terms of Gamma functions:
\begin{equation}
\begin{aligned}
\label{eqn:gamma-coefficients}
A^+_{\w\ell}(\Delta) &= \frac{N^{\Delta - N(|\ell| + \w) - 2}\, \G(\Delta - 1) \G(1 + N|\ell|)}{\G \Big( \frac{1}{2} \big( \Delta + N(|\ell| + \w) \big) \! \Big) \G \Big( \frac{1}{2} \big( \Delta + N(|\ell| - \w) \big) \!\Big)}, \\ 
A^-_{\w\ell}(\Delta) &= A^+_{\w\ell}(2 - \Delta).
\end{aligned}
\end{equation}
Unless $A^+_{\w\ell}(\Delta)$ vanishes, the $r^{\Delta - 2}$ term in (\ref{eqn:asymptotics}) will cause $R_{\w\ell}(r)$ to diverge at infinity.\footnote{This is only true when $\Delta > 2$. A scalar of mass $m$ in $\mathrm{AdS}_3$ is subject to the Breitenlohner--Freedman bound $m^2 \geq -1$ and can have one of two scaling dimensions: $\Delta = \Delta_{\pm} = 1 \pm \sqrt{1 + m^2}$ \cite{Breitenlohner:1982jf}. When $m^2 > 0$, only $\Delta = \Delta_+ > 2$ leads to a consistent theory, precisely because neither term in (\ref{eqn:asymptotics}) would be normalizable if we had used $\Delta = \Delta_-$. But in the range $-1 < m^2 < 0$, both $\Delta_+ \in (1,2)$ and $\Delta_- \in (0,1)$ produce normalizable modes, albeit with different choices of boundary conditions. Here and henceforth we adopt $\Delta = \Delta_+$, but we will return to this issue briefly in \S\ref{sec:thermal-physics} from a different perspective.} Consequently, we require $A^+_{\w\ell}(\Delta) = 0$: this condition is satisfied at the poles of the two Gamma functions in the denominator of $A^+_{\w\ell}(\Delta)$, which occur when their arguments assume nonpositive integer values. Therefore $R_{\w\ell}(r)$ is only normalizable if $\w$ is quantized:
\begin{align}
\label{eqn:quantization}
A^+_{\w\ell}(\Delta) = 0 \iff
\w = \w_{n\ell} \equiv \pm \frac{1}{N}\big(\Delta + 2n + N|\ell|\big), \qquad n \in \N.
\end{align}
In order to uniquely define the vacuum state of the theory once it is quantized, we choose the positive sign above, so that $\w > 0$. This is consistent with pure $\mathrm{AdS}_3$ ($N=1$), where the descendants of $\phi$, i.e. the operator insertions $\Box^n \partial^\ell \phi$ in the dual field theory, will have energy $\Delta+2n+|\ell|$ on the cylinder. We relabel all of the modes by $n$ and $\ell$, and substitute (\ref{eqn:quantization}) into (\ref{eqn:radial-solutions}) to see that the normalizable radial modes take the form
\begin{equation}
\begin{aligned}
\label{eqn:radial-modes}
R_{n\ell}(r) &= \mathcal{N}_{n\ell} \p{r^2 + \frac{1}{N^2}}^{\frac{1}{2} \p{\Delta + 2n + N|\ell|}} r^{N|\ell|} \mathcal{F}_{n\ell}(r), \\
\mathcal{F}_{n\ell}(r) &= {}_2 F_1 \Big[ 1 + n + N|\ell|,\, \Delta + n + N|\ell|;\, 1+N|\ell|;\, -N^2 r^2 \Big].
\end{aligned}
\end{equation}

Having ensured that the integral (\ref{eqn:mode-norm}) is finite, we evaluate it by direct substitution of (\ref{eqn:radial-modes}) in Appendix \ref{sec:normalization}. The resulting value of $\mathcal{N}_{n\ell}$ that normalizes the modes is
\begin{align}
\label{eqn:normalization-constant}
\mathcal{N}_{n\ell}^2 = \p{\frac{\G(1 + n + N|\ell|)\, \G(\Delta + n + N|\ell|)}{2\pi n!\, \G(1 + N|\ell|)^2\, \G(n + \Delta)}} N^{2\Delta + 4\p{n + N|\ell|} + 1}.
\end{align}

\subsection{The Feynman Propagator}
\label{sec:bdy-wightman}

We proceed to quantize the theory. The field is promoted to an operator via (\ref{eqn:commutators}):
\begin{align}
\label{eqn:full-solution}
\phi(x) = \sum_{\ell \in \Z} \sum_{n \in \N} \Big( a_{n\ell} e^{-i\w_{n\ell} t} e^{i\ell\th} + a_{n\ell}^{\dagger} e^{i\w_{n\ell} t} e^{-i\ell\th} \Big) R_{n\ell}(r), \quad \big[a_{n\ell},\, a_{n'\ell'}^{\dag}\big] = \d_{nn'} \d_{\ell\ell'}.
\end{align}
The choice $\w > 0$ in (\ref{eqn:quantization}) is crucial: the sign of $\w$ in (\ref{eqn:full-solution}) determines whether $a_{n\ell}$ or $a^{\dagger}_{n\ell}$ function as the annihilation operators that destroy the vacuum of the QFT on the conical background. If we want the vacuum to have minimal energy, all operators that add energy must be raising operators, and the lowering operators must annihilate the vacuum.

The bulk-to-bulk Wightman function is then given by (\ref{eqn:wightman0}):
\begin{align}
G^+_{N}(x_1,x_2) = \sum_{\ell \in \Z} \sum_{n \in \N} e^{-i\w_{n\ell}\p{t_1 - t_2}} e^{i\ell\p{\th_1 - \th_2}} R_{n\ell}(r_1) R_{n\ell}(r_2).
\end{align}
Let us evaluate $G^+_{N}(x_1,x_2)$ in the limit $r,r' \too \infty$, as a function of the boundary opening angle $\th = \th_1 - \th_2$ and the time delay $t = t_1 - t_2$. Near the boundary, the radial modes---now normalizable---decay as $R_{n\ell}(r) \sim \mathcal{N}_{n\ell} A^-_{n\ell}(\Delta) r^{-\Delta}$, thanks to (\ref{eqn:asymptotics}). According to the extrapolate dictionary \cite{Witten:1998qj,Gubser:1998bc}, the boundary Wightman function is given by
\begin{align}
\label{eqn:wightman1}
G^+_N(t,\th) = \lim_{r_1,r_2 \to \infty} \! \Big( r_1^{\Delta} r_2^{\Delta} G^+_{N}(x_1,x_2) \Big) = \sum_{\ell \in \Z} \sum_{n \in \N} e^{-i\w_{n\ell} t} e^{i\ell\th} \Big( \mathcal{N}_{n\ell} A^-_{n\ell}(\Delta) \Big)^2.
\end{align}
Combining (\ref{eqn:gamma-coefficients}) with the normalization constant (\ref{eqn:normalization-constant}) produces a few cancellations:
\begin{equation}
\label{eqn:fourier-coefficients}
\begin{aligned}
\mathcal{C}_{n\ell}^2 \equiv \Big( \mathcal{N}_{n\ell} A^-_{n\ell} \Big)^2 &= \frac{N^{1-2\Delta}}{2\pi \G(\Delta)^2} \frac{\G(\Delta + n) \G(\Delta + n + N|\ell|)}{\G(1+n) \G(1 + n + N|\ell|)}.
\end{aligned}
\end{equation}
The sum over $n$ in (\ref{eqn:wightman1}), with Fourier coefficients (\ref{eqn:fourier-coefficients}) and $\w_{n\ell}$ given by (\ref{eqn:quantization}), can be performed in \textit{Mathematica}. The Wightman function reduces to a single sum over $\ell$:
\begin{equation}
\label{eqn:wightman2}
\begin{aligned}
G^+_N(t,\th) &= \sum_{\ell \in \Z} \sum_{n \in \N} \mathcal{C}_{n\ell}^2 e^{-i\w_{n\ell} t} e^{i\ell\th} = \frac{N^{1 - 2\Delta}}{2\pi \G(\Delta)} \sum_{\ell \in \Z} s_{\ell}(t) e^{-\frac{i}{N}(\Delta + N|\ell|) t}e^{i\ell\th}, \\
s_{\ell}(t) &= \p{\frac{\G(\Delta + N|\ell|)}{\G(1 + N|\ell|)}}\, {}_2 F_1 \Big[\Delta,\, \Delta + N|\ell|;\, 1 + N|\ell|;\, e^{-\frac{2it}{N}}\Big].
\end{aligned}
\end{equation}

We split the sum over $\ell$ into positive and negative pieces (plus a zero mode), relabel $\ell \too -\ell$ in the latter sum, and express (\ref{eqn:wightman2}) in light-cone coordinates $x^{\pm} = t \pm \th$:
\begin{align}
\label{eqn:wightman3}
G^+_N(t,\th) &= \frac{N^{1-2\Delta}}{2\pi} e^{-\frac{i\Delta t}{N}} {}_2 F_1 \Big[\Delta, \Delta; 1; e^{-\frac{2it}{N}}\Big] + \frac{N^{1 - 2\Delta}}{2\pi \G(\Delta)} e^{-\frac{2i\Delta t}{N}} \sum_{\ell = 1}^{\infty} s_{\ell}(t) \p{e^{-i\ell x^+} + e^{-i\ell x^-}} = \notag \\
&= \frac{N^{1-2\Delta}}{2\pi} e^{-\frac{i\Delta t}{N}} {}_2 F_1 \Big[\Delta, \Delta; 1; e^{-\frac{2it}{N}}\Big] + \frac{N^{1 - 2\Delta}}{\pi \G(\Delta)} \, e^{-\frac{2i\Delta t}{N}}\,  \sum_{\ell = 1}^{\infty} s_{\ell}(t) e^{-i\ell t} \cos\p{\ell \th}.
\end{align}
The fact that $s_{\ell}(t)$ grows polynomially with $\ell$ prevents the series above from converging. To improve the situation, one may shift $t \too (1 - i\eps) t$ slightly off the real axis. This $i\eps$ prescription introduces exponentially decaying factors in the phases $e^{-i\ell t}$ above, regulates the series, and allows the limit $\eps \too 0$ to be taken safely. 

The Feynman propagator $G_N(t,\th)$ is constructed from the Wightman functions $G_N^{\pm}(t,\th)$:
\begin{equation}
\label{eqn:heaviside}
\begin{aligned}
G_N(t,\th) &= \Theta(t_1 - t_2) G_N^+(t,\th) + \Theta(t_2 - t_1) G_N^-(t,\th) = \\ &= \Theta(t) G_N^+(t,\th) + \Theta(-t) G_N^+(-t,\th).
\end{aligned}
\end{equation}
Here $\Theta$ is the Heaviside step function, with $\Theta(0) = \frac{1}{2}$ by convention, and we have used the fact that $G_N^+$ is even in $\th$ to conclude that $G_N^-(t,\th) = G_N^+(-t,-\th) = G_N^+(-t,\th)$.

\begin{figure}[t]
    \centering
    \includegraphics[width=.9\textwidth]{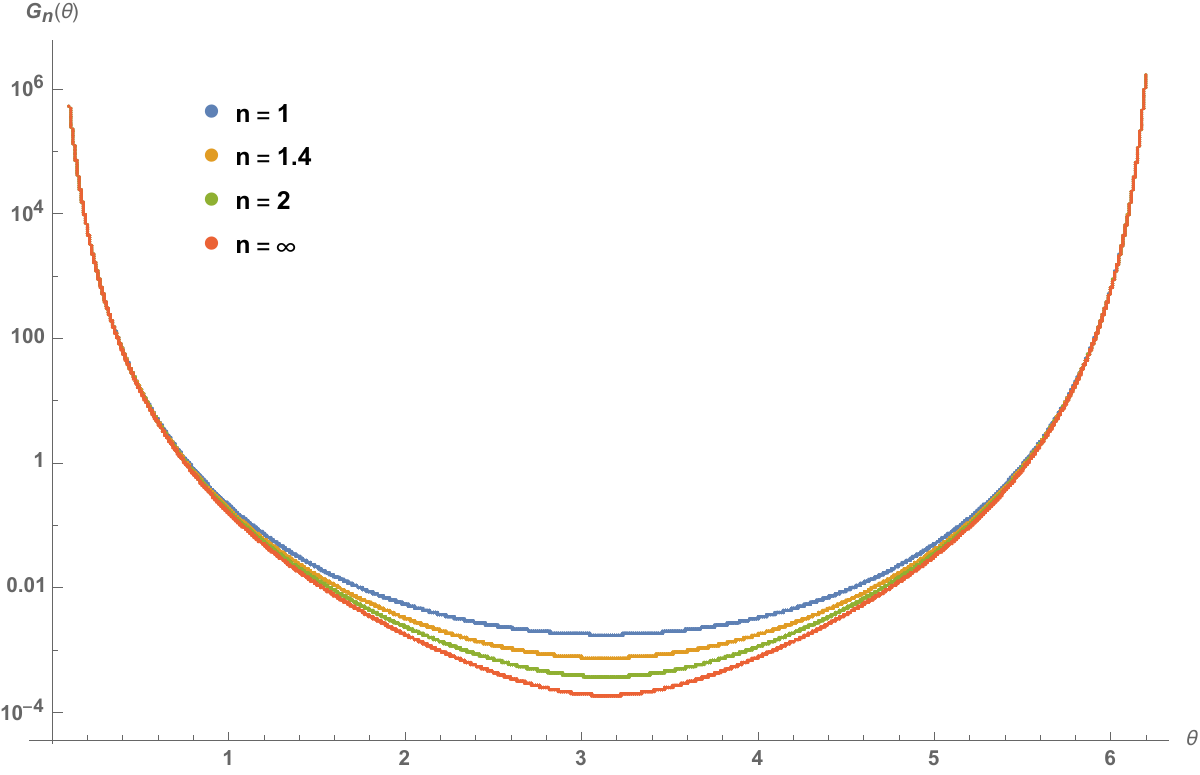}
    \caption{The equal-time propagator $G_N(\th)$ is plotted for $\Delta = 3.25$ and several values of $N$. Correlations decay as $N$ increases, especially for large separations near $\th = \pi$.}
    \label{fig:propagators}
\end{figure}

At equal times ($t = 0$), the propagator simplifies: the hypergeometric functions in (\ref{eqn:wightman2}) reduce at unit argument to ratios of Gamma functions,\footnote{The hypergeometric series for $s_{\ell}(0)$ in (\ref{eqn:wightman2}) converges only for $\Delta \in (0, \frac{1}{2})$. We evaluate it in this range, observe that the result is analytic in $\Delta$, and analytically continue to obtain the result for $\Delta \geq \frac{1}{2}$.} and the phases $e^{-i\p{\Delta + N|\ell|} t/N}$ disappear. From (\ref{eqn:heaviside}) we have $G_N(0) = G_N^+(0, \th)$; when the dust settles, we find
\begin{align}
\label{eqn:main-result}
G_N(\th) = \k_N \sum_{\ell \in \Z} s_{\ell} e^{i\ell\th}, \qquad 
s_{\ell} = \frac{\G(\Delta + N|\ell|)}{\G(1 + N|\ell| - \Delta)}, \qquad 
\k_N = \frac{N^{1 - 2\Delta}}{4\pi \cos\p{\pi \Delta} \G(2\Delta)}.
\end{align}
This result is illustrated in Fig. \ref{fig:propagators}. When $\Delta$ is a half-integer, the prefactor $\k_N$ diverges; in this case, the modified propagator $\tilde{G}_N(\th) = \cos\p{\pi \Delta} G_N(\th)$ is non-singular.

\section{The Structure of the Propagator}
\label{sec:properties}

Here we discuss in detail the behavior of the propagator. We begin with its light-cone singularities, and then specialize to the case of equal times to analyze its UV behavior. Then we consider the quotients $\mathrm{AdS}_3/\Z_N$, where the method of images gives a closed-form expression for the propagator and allows us to interpret in terms of bulk geodesics.

\subsection{Light-Cone Singularities}
\label{sec:light-cone}

It is well known that propagators usually diverge at null separation \cite{Friedlander:abc}. In our case, the light cones on the boundary are defined by $x^{\pm} = t \pm \th = 0$; we seek the singularities of $G_N(t,\th)$ there. Taking $x^+ = 0$ in (\ref{eqn:wightman3}) leaves the sum of the coefficients $s_{\ell}(t)$ exposed, and the hypergeometric functions in $s_\ell(t)$ diverge when their argument $e^{-2i t/N}$ is equal to one. (Otherwise, $e^{-2it/N}$ lands in the unit circle away from 1, where the ${}_2 F_1$ is well behaved.) So $G_N^+(t,\th)$ and $G_N(t,\th)$ can be singular when $t = N\pi k$ for $k \in \Z$. The same holds for $x^- = 0$, except that now it is the $e^{i\ell x^-}$ terms which can cause $G_N^+(t,\th)$ to diverge. In fact, more is true: whenever $x^{\pm} \in 2\pi \Z$, the light cones wrap the boundary of the $\mathrm{AdS}$ cylinder, and $G_N(t,\th)$ can diverge again for certain values of $t$ and $\th$.

A more direct way to see this follows from (\ref{eqn:wightman1}), which diverges when all of the phase factors $e^{-i\w_{n\ell} t} e^{i\ell\th}$ are mutually equal. If two such phases labeled by $(n,\ell)$ and $(n',\ell')$ have $\ell = \ell'$, then they match when $e^{-i\w_{n\ell} t} = e^{-i\w_{n'\ell} t}$. This occurs at times $t_{n,n'}$ given by
\begin{align}
\p{\w_{n\ell} - \w_{n'\ell}}t_{n,n'} = \frac{2}{N} \p{n - n'} t_{n,n'} = 2\pi k \iff t_{n,n'} = \frac{Nk\pi}{n-n'}, \qquad k \in \Z.
\end{align}
In particular, $t_k = Nk\pi$ is an integer multiple of all possible pairwise resonances $t_{n,n'}$, since $n-n' \in \Z$. So if $G_N(t,\th)$ diverges, then it must do so at $t = t_k$. Next, we seek the locus of positions $\th$ where $G_N(t_k, \th)$ diverges: this occurs when all phases, including those with $\ell \neq \ell'$, line up at $t = t_k$. Therefore once again we consider two phases $(n,\ell)$ and $(n', \ell')$. These interfere constructively at positions $\th_{\ell,\ell'}$ given by
\begin{align}
\p{\w_{n\ell} - \w_{n'\ell'}} t_k - \p{\ell - \ell'}\th_{\ell,\ell'} = 2\pi k' \iff \th_{\ell,\ell'} = \frac{\p{|\ell| - |\ell'|} t_k - 2\pi k'}{\ell - \ell'}, \quad k' \in \Z.
\end{align}
Once again, taking $\ell$ and $\ell'$ to be neighboring integers gives the condition under which all phases are mutually equal. We find, after being careful with the absolute values, that
\begin{align}
\label{eqn:light-cone}
\th_{k,k'} = \pm t_k - 2\pi k' = \pm Nk\pi - 2\pi k' \iff t_k \pm \th_{k,k'} = x^{\pm}_{k,k'} \in 2\pi \Z.
\end{align}
When $N$ is a positive integer, the propagator diverges at vanishing angular separation---a universal feature that we discuss below---and, if $N$ is odd, also at $\th = \pi$. In pure $\mathrm{AdS}_3$ ($N=1$), the first non-trivial divergence occurs at the opposite side of the cylinder from where we start, while traveling at the speed of light on the boundary, as expected.

\subsection{Short-Distance Behavior}
\label{sec:universal-UV}

When the separation between two boundary points is small, $G_N(\th)$ becomes singular due to the polynomial growth of its Fourier coefficients in $\ell$. By power-counting, we expect a divergence of the form $G_N(\th) \sim \th^{-2\Delta}$. This matches the universal form of such UV divergences in asymptotically AdS spacetimes \cite{Witten:1998qj,Gubser:1998bc}, and is the standard divergence of OPEs that arises when CFT operator insertions on the boundary approach each other. We can better understand the divergence by studying the asymptotics of the coefficients $s_{\ell}$ at large $\ell$. The Stirling series for a ratio of Gamma functions \cite{F.TRICOMI} begins
\begin{equation}
\begin{aligned}
\label{eqn:sl-asymptotics}
s_{\ell} &= \frac{\G(\Delta + N|\ell|)}{\G(1 + N|\ell| - \Delta)} \sim \p{N|\ell|}^{2\Delta - 1} - \frac{\Delta}{12} \p{2\Delta - 1}\p{2\Delta - 2} \p{N|\ell|}^{2\Delta - 3} \,+ \\ &+
\frac{\Delta}{360}\p{\Delta-2} \p{\Delta-1} \p{2\Delta-3}\p{2\Delta-1}\p{5\Delta+1} \p{N|\ell|}^{2\Delta - 5} + O\p{N|\ell|^{2\Delta - 7}}.
\end{aligned}
\end{equation}

Since the singular structure of $G_N(\th)$ is controlled by the behavior of its large-$\ell$ terms, we can replace the Fourier coefficients in (\ref{eqn:main-result}) by the asymptotic expansion above, and sum the Fourier series for each term in the expansion. Doing so---one gets polylogarithms in $e^{i\th}$---and making a power expansion in $\th$, we find the following singular behavior:
\begin{equation}
\label{eqn:bulk-expansion}
\begin{aligned}
G_N(\th) &=
\frac{\th^{-2\Delta}}{2\pi} \p{1 + \frac{\Delta}{12 N^2} \th^2 + \frac{\Delta \p{5\Delta + 1}}{1440N^4} \th^4 + \frac{\Delta\p{4 + 7\Delta \p{3 + 5\Delta}}}{362880 N^6} \th^6 + O\p{\th^8}}.
\end{aligned}
\end{equation}
This confirms that $G_N(\th)$ is normalized and has the correct leading UV divergence.

As a bonus, we observe that $\ell$ appears in (\ref{eqn:sl-asymptotics}) only through the combination $N|\ell|$. This makes (\ref{eqn:bulk-expansion}) also the expansion of $G_N(\th)$ at large $N$. As $N \too \infty$ asymptotically, we retain only the leading term $s_{\ell} = \p{N|\ell|}^{2\Delta - 1}$. There is then a cancellation between the powers of $N$ between the $N^{2\Delta - 1}$ inside the sum and the contribution of $\k_N$, and we find that the scalar propagator in the massless BTZ spacetime is
\begin{align}
\label{eqn:mBTZ}
G_{\infty}(\th) = \k_N \sum_{\ell \in \Z} \p{N|\ell|}^{2\Delta - 1} e^{i\ell\th} = \frac{\mathrm{Li}_{1 - 2\Delta}(e^{i\th}) + \mathrm{Li}_{1 - 2\Delta}(e^{-i\th})}{4\pi \cos\p{\pi \Delta} \G(2\Delta)}.
\end{align}

\subsection{The Method of Images}
\label{sec:method-of-images}

\begin{figure}[t]
    \centering
    \includegraphics[width=.9\textwidth, trim={95 290 105 45}, clip]{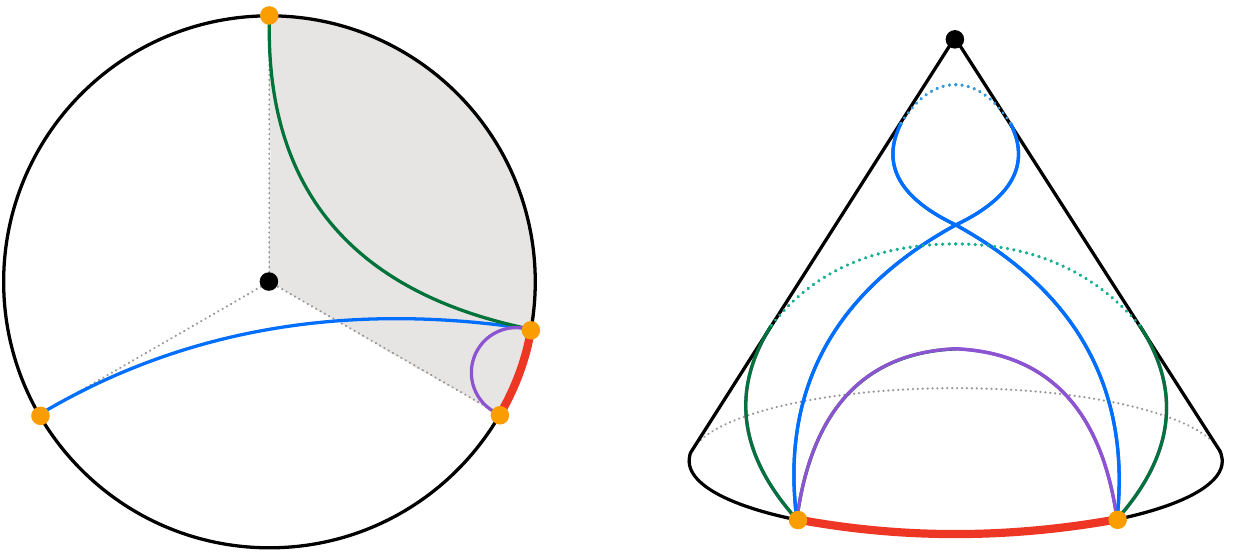}
    \caption{Left: a constant-time slice of $\mathrm{AdS}_3$, with the fundamental domain for a $\Z_N$ action shaded. Geodesics between one boundary endpoint and the images of another are shown. Right: the geodesics of the cover descend to distinct geodesics between the same endpoints in the $\mathrm{AdS}_3/\Z_N$ quotient. All but two wind nontrivially around the conical singularity and are never minimal.}
    \label{fig:winding}
\end{figure}

When $N$ is an integer, the method of images for Green's functions gives a closed form for the propagator \cite{Arefeva:2014aoe}. The idea is to pass to the universal cover of $\mathrm{AdS}_3/\Z_N$, which is $\mathrm{AdS}_3$ itself. Points of $\mathrm{AdS}_3$ with angular coordinates differing by $\frac{2\pi k}{N}$, for $k \in \{0, ..., N-1\}$, are identified in the quotient (see Fig. \ref{fig:winding}), and the scalar's propagation from $(0,0)$ to $(t,\th)$ in $\mathrm{AdS}_3/\Z_N$ is equivalent to its propagation from $(0,0)$ to all of the images $(t,\th + \frac{2\pi k}{N})$ in the cover. For proof, consider the Fourier series (\ref{eqn:main-result}). When $N \in \Z_+$, we can drop the absolute values in the coefficients $s_{\ell}$.\footnote{\textit{Proof.} It suffices to show that removing the absolute values in $s_{\ell}$ leaves it even in $\ell$ when $N \in \Z_+$. This follows from the reflection formula $\G(z) \G(1-z) = \frac{\pi}{\sin\p{\pi z}}$: $s_{\ell} = \frac{\Gamma(N\ell + \Delta)}{\Gamma(1 - \Delta + N\ell)} = \frac{\Gamma(\Delta - N\ell)}{\Gamma(1-\Delta - N\ell)} \cdot \frac{\sin\qty[\pi\p{\Delta - N\ell}]}{\sin\qty[\pi\p{\Delta + N\ell}]} = s_{-\ell}$. The hypothesis $N \in \Z_+$ was necessary in the last step, in order to show that $\frac{\sin\qty[\pi\p{\Delta - N\ell}]}{\sin\qty[\pi\p{\Delta + N\ell}]} = 1$. \qed} To exploit the fact that the $s_{\ell}$ depend on $\ell$ only through the product $N\ell$, we define $\s_{N\ell} \equiv s_{\ell}$. We then rescale $\ell \too \ell' \equiv N\ell$:
\begin{align}
\label{eqn:images-proof}
G_N(\th) &= \k_N \sum_{\ell \in \Z} \s_{N\ell} e^{i\ell\th} = \k_N \! \sum_{\ell' \in N\Z} \s_{\ell'} e^{i\ell'\th/N} = \k_N \sum_{\ell' \in \Z} \s_{\ell'} e^{i\ell'\th/N} \bigg(\frac{1}{N} \sum_{k = 0}^{N-1} e^{2\pi i k \ell'/N}\bigg) = \notag \\ &= \frac{\k_N}{N} \sum_{k = 0}^{N-1} \sum_{\ell' \in \Z} \s_{\ell'} e^{i\ell'\p{\th + 2\pi k}/N} = N^{-2\Delta} \sum_{k = 0}^{N-1} G_1\p{\frac{\th + 2\pi k}{N}}.
\end{align}
The rescaling is effectively a coordinate redefinition that ensures that $\th \sim \th + 2\pi$ rather than $\th \sim \th + \frac{2\pi}{N}$ in the quotient space. This factor of $N$ is what explains how angular momentum is quantized. In any case, it remains only to compute $G_1$ explicitly.

But the form of $G_1(t,\th)$ follows from our main results (\ref{eqn:wightman2}) and (\ref{eqn:main-result}) at $N=1$:\footnote{\textit{Proof.} We recognize (\ref{eqn:main-result}) at $N=1$ as a sum of two ${}_2 F_1$ functions in the variables $e^{\pm i\th}$, which can be expressed in terms of $x \equiv \sin\p{\frac{\th}{2}}$ as $e^{\pm i\th} = 1 - 2x^2 \pm i \sin\qty[2 \sin^{-1}(x)]$. We then expand the resulting hypergeometric series in powers of $x$. We find, besides the expected term $\frac{x^{-2\Delta}}{2\pi}$, that the next few terms all vanish. The coefficients of hypergeometric functions are defined recursively, so this observation shows that the series for $G_1(\th)$ terminates after $\frac{x^{-2\Delta}}{2\pi} = \frac{1}{2\pi} \big[2\sin(\th/2)\big]^{-2\Delta}$. The argument for $t \neq 0$ is similar. \qed}
\begin{align}
\label{eqn:n=1}
G_1(t,\th) = \frac{1}{2\pi} \Big[2\big(\!\cos t - \cos\th \big)\Big]^{-\Delta} \implies G_1(\th) = \frac{1}{2\pi} \qty[2 \sin\p{\frac{\th}{2}}]^{-2\Delta}.
\end{align}
Finally, we construct $G_N(\th)$ by substituting (\ref{eqn:n=1}) into  (\ref{eqn:images-proof}):
\begin{align}
\label{eqn:images-result}
G_N(\th) = \sum_{k=0}^{N-1} N^{-2\Delta} G_1\p{\frac{\th + 2\pi k}{N}} = \frac{1}{2\pi} \sum_{k=0}^{N-1} \qty[2N \sin\p{\frac{\th + 2\pi k}{2N}}]^{-2\Delta}.
\end{align}
To see that this propagator is properly normalized, note that its UV divergence is due solely to its $k=0$ term $N^{-2\Delta} G_1\p{\frac{\th}{N}} \sim \frac{1}{2\pi} \th^{-2\Delta}$, which matches (\ref{eqn:bulk-expansion}). The same reasoning above goes through without change for the unequal-time propagator $G_N(t,\th)$:
\begin{align}
\label{eqn:images-time}
G_N(t,\th) = \sum_{k=0}^{N-1} N^{-2\Delta} G_1\big(\tfrac{t}{N},\, \tfrac{\th + 2\pi k}{N}\big) = \frac{1}{2\pi} \sum_{k=0}^{N-1} \bigg[2N^2\Big(\! \cos\p{\tfrac{t}{N}} - \cos\p{\tfrac{\th + 2\pi k}{N}}\Big)\bigg]^{-\Delta}.
\end{align}
The factors of $N$ appearing in (\ref{eqn:images-time}) may seem strange, but recall that the $(t,\theta)$ coordinates on the covering space are different than the ones we started with: they differ by a factor of $N$. We have seen how this affected angles; it also extends to time. The prefactors of $N^{-2\Delta}$ in (\ref{eqn:images-result}--\ref{eqn:images-time}) arise from the proper conformal rescaling of the two-point functions between both coordinate systems at the boundary; each $\phi$ gets rescaled by $N^{-\Delta}$.

\subsection{The Geodesic Approximation}
\label{sec:geodesic-approximation}

The propagator derived above has a nice interpretation in the context of the geodesic method for two-point functions \cite{Balasubramanian:1999zv}. In the semiclassical limit of a worldline path integral over paths $\mathcal{P}$, the propagator $G_N(x,x')$ localizes to a saddle-point sum over all geodesics $\g_k(x,x')$ between $x$ and $x'$, weighted by their renormalized lengths $\L[\g_k(x,x')]$:
\begin{align}
\label{eqn:geodesic-saddles}
G_N(x,x') = \big\langle \phi(x) \phi(x') \big\rangle_{N} \propto \int_x^{x'} \mathcal{D} \mathcal{P}\, e^{-m \int \dd s} \approx \sum_{k} e^{-m \L[\g_k(x,x')]}.
\end{align}

In pure $\mathrm{AdS}_3$, a unique geodesic connects any two points $p \neq q$ \cite{Ballmann}, so there are $N$ geodesics between $p$ and the $\Z_N$ images $q_k$ of $q$. The $q_k$ are all identified in the quotient, but the $\mathrm{AdS}_3$ geodesics descend to distinct geodesics $\g_0, ..., \g_{N-1}$ in $\mathrm{AdS}_3/\Z_N$ between the same two boundary points. As shown in Fig. \ref{fig:winding}, some of the $\g_k$ wind around the cone and attain a winding number $w_k$ given by the number of $\Z_N$ sectors they cross in the $\mathrm{AdS}_3$ cover. We compute the lengths $\L[\g_k(t,\th)]$ of spacelike geodesics in $\mathrm{AdS}_3/\Z_N$ in Appendix \ref{sec:geodesic-length}. For purely spatial geodesics ($t = 0$), our results agree with those of \cite{Balasubramanian:2014sra}:
\begin{align}
\label{eqn:geodesic-length}
\L[\g_k(\th)] = 2 \ln\qty[2N \sin\p{\frac{\a_k}{2N}}] = 2\ln\qty[2N \sin\p{\frac{\th + 2\pi k}{2N}}].
\end{align}
Here $\a_k = \min\{\th,\, 2\pi - \th\} + 2\pi w_k$ is the boundary opening angle subtended by $\g_k$.\footnote{Our definition of $\a_k$ differs from that of \cite{Balasubramanian:2014sra} by a factor of two. Together with $w_k = \min\{k,\, N-k-1\}$, we have $\a_k = \min\{\th + 2\pi k,\, 2\pi N - (\th + 2\pi k)\}$, which proves the second equality in (\ref{eqn:geodesic-length}).}

When we substitute the geodesic length (\ref{eqn:geodesic-length}) into the saddle-point sum (\ref{eqn:geodesic-saddles}), the result matches the image formula (\ref{eqn:images-result}) up to a normalization constant and in the limit of a heavy scalar, where $\Delta = 1 + \sqrt{1 + m^2} \approx m$. A more precise calculation that includes zero-point energy corrections, as in \cite{Berenstein:2002ke}, would show that it is really $\Delta$ that shows up in the exponent. An analogous match can be made at unequal times, where instead
\begin{align}
\label{eqn:geodesic-length-time}
\L[\g_k(t,\th)] = \ln \qty[2N^2 \p{\cos \p{\frac{t}{N}} - \cos\p{\frac{\th + 2\pi k}{N}}}].
\end{align}
This produces the propagator (\ref{eqn:images-time}) upon substitution into the sum over geodesics. See Appendix \ref{sec:geodesic-length} for an explicit calculation of the geodesic lengths (\ref{eqn:geodesic-length}) and (\ref{eqn:geodesic-length-time}). 

The geodesic approximation gives a geometric interpretation to the sum over images: it emphasizes that contributions to $G_N(\th)$ from geodesics that wrap the singularity are suppressed, while the dominant contribution comes from the minimal geodesic. In particular, in Fig. \ref{fig:winding} we see that when $\th$ crosses $\pi$, the two geodesics on either side of the defect---$\g_0$ and $\g_{N-1}$---attain comparable length and exchange dominance. This exchange is an essential characteristic of the Ryu--Takayanagi phase transition: see \S\ref{sec:RT}.

Previous work \cite{Arefeva:2014aoe, Ageev:2015qbz, Arefeva:2016nic, Arefeva:2016wek} has attempted to extend the geodesic method to the case of non-integer $N$, but the resulting propagator is discontinuous in $\th$. This is unfortunate but not surprising, since the number of geodesics between two points in conical $\mathrm{AdS}_3$ is not uniform throughout the spacetime: if $N \notin \Z_+$, it varies discontinuously with the separation between the points. This is one example where saddle-point path integral techniques cannot be applied na\"ively, even for free theories on a fixed background. One must sometimes resort to the canonical formalism, as we have done in \S\ref{sec:scalar-propagator}, to gain full control.

It is natural to ask whether the results obtained from the method of images for integer $N$ can be analytically continued to arbitrary $N \in \R_+$, obviating the need for the analysis of \S\ref{sec:scalar-propagator}. One immediate difficulty is that (\ref{eqn:images-result}) and (\ref{eqn:images-time}) are both sums of $N$ terms, a notion that only makes sense for $N \in \Z_+$. Nevertheless, section 6 of \cite{Calabrese:2010he} gives a clever way to proceed: one writes each of the $N$ summands of (\ref{eqn:images-result}) as Fourier series whose coefficients can be computed explicitly, and then uses Carlson's theorem \cite{Hardy:abc} to control the behavior of the resulting analytic function at infinity. This argument reproduces the Fourier series (\ref{eqn:main-result}) exactly and provides a shortcut to our main result; however, it also obscures the story of the propagator's origins. We will now give the CFT perspective on this story.

\section{Heavy-Light Correlators in CFT}
\label{sec:CFT-interpretation}

In this section, we interpret our results from the perspective of the dual CFT, where the propagator is a four-point function involving the scalar and the defect. Our first step will be to recast the bulk scalar's excitations on top of the defect as two-particle states in the CFT. We will show that the contributions of these states to the correlator are encoded in the mode sum of \S\ref{sec:scalar-propagator}, which takes the form of an s-channel OPE. Meanwhile, the t-channel OPE is dominated by the Virasoro vacuum block, which is particularly simple in the semiclassical limit. We use it to match the singular structure of the s channel, and to better understand the method of images and the ``thermal'' nature of the correlator.

\subsection{OPEs and Multiparticle States}

\begin{figure}[t]
    \centering
    \scalebox{0.75}{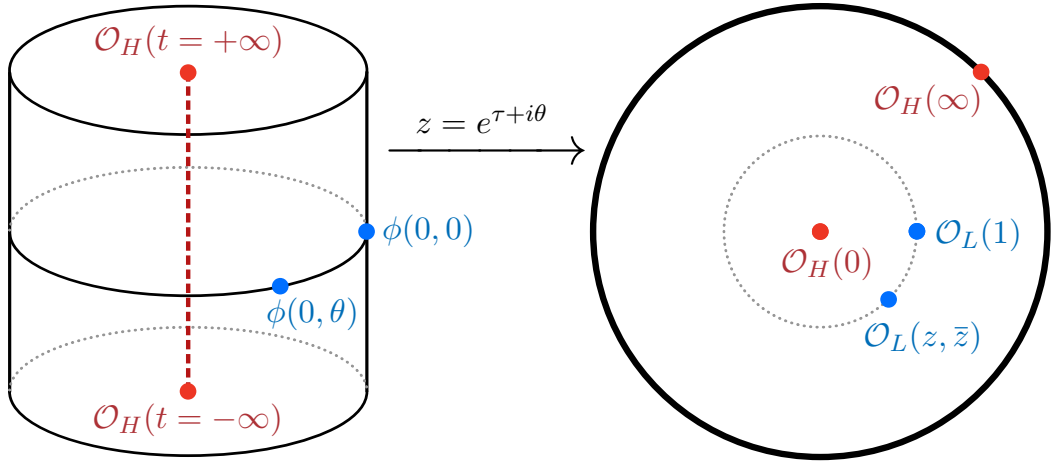}
    \vspace{-11pt}
    \caption{This figure illustrates our setup from the perspectives of the conical bulk geometry and the dual CFT on the plane. We are particularly interested in the limit of nearby $\mathcal{O}_L$ insertions.}
    \label{fig:radial-quantization}
\end{figure}

In radial quantization, we conformally map the AdS boundary cylinder to the punctured plane. We adopt complex coordinates $z = e^{\t + i\th}$ and $\bar{z} = e^{\t - i\th}$, where $\t = it$ is the Euclidean time and $\th \sim \th + 2\pi$. The bulk field $\phi$ is dual to a light scalar primary operator $\mathcal{O}_L$ with conformal dimension $\Delta_L = \Delta$, while the conical defect is dual to a heavy scalar primary operator $\mathcal{O}_H$ whose conformal dimension $\Delta_H \gg \Delta_L$ is given by
\begin{align}
\Delta_H = M = \frac{1}{8G} \p{1 - \frac{1}{N^2}} = \frac{c}{12} \p{1 - \a^2}, \qquad
\a \equiv \frac{1}{N} = \sqrt{1 - \frac{12 \Delta_H}{c}}.
\end{align}
Here we have used the relation $c = \frac{3}{2G}$ \cite{Brown:1986nw} and introduced the defect parameter $\a \in (0,1]$. By the operator-state correspondence, the defect operator defines an excited state via $\ket{\mathcal{O}_H} = \mathcal{O}_H(0) \ket{0}$. This state is dual to the conical background geometry: see Fig. \ref{fig:radial-quantization}.

In this setting, the scalar propagator is a normalized 
semiclassical heavy-heavy-light-light (HHLL) four-point function involving both $\mathcal{O}_L$ and $\mathcal{O}_H$:
\begin{align}
\label{eqn:correlator-def}
G_N(z, \bar{z}) = 
\frac{\big\langle \mathcal{O}_H(\infty) \mathcal{O}_L(1) \mathcal{O}_L(z, \bar{z}) \mathcal{O}_H(0)\big\rangle }{\big\langle \mathcal{O}_H(\infty) \mathcal{O}_H(0)\big\rangle}.
\end{align}
This correlator can be computed in one of two OPE channels, each of which sums over the Virasoro conformal blocks of all local primary operators in the theory.
\begin{itemize}
\item \textbf{The t-channel (``HHLL'') OPE} converges in $|1 - z| < 1$, i.e. when the two light operators are brought together. By dimensional analysis, this OPE is dominated by the Virasoro block of the lightest primary in the theory---the identity \cite{Fitzpatrick:2014vua}.
\item \textbf{The s-channel (``HLLH'') OPE} converges in $|z| < 1$, i.e. when $\mathcal{O}_L(z,\bar{z})$ is brought close to $\mathcal{O}_H(0)$. As shown in \cite{Fitzpatrick:2014vua}, this OPE is dominated by the exchange of an infinite family of ``double-twist'' primaries, which we describe presently.
\end{itemize}
The double-twist primaries $\mathcal{O}_{n\ell} = [\mathcal{O}_H \mathcal{O}_L]_{n\ell}$ are composite operators labeled by excitation number $n \in \N$ and angular momentum $\ell \in \Z$. They describe two-particle states in $\mathrm{AdS}_3$ consisting of Fourier modes of the scalar orbiting the defect with a given energy and angular momentum. (See \cite{Berenstein:2021pxv, Berenstein:2014cia, Berenstein:2019tcs} for arguments along similar lines.) Their spectrum \cite{Fitzpatrick:2014vua} is\footnote{More precisely, the CFT spectrum is required to contain families of primaries in ``Regge trajectories'' whose spectrum is asymptotic to (\ref{eqn:spectrum}) at large $\ell$ \cite{Collier:2018exn}. But the large-spin asymptotics will be sufficient for our purposes, since the modes with $\ell \gg 1$ control the singular structure of the correlator.}
\begin{align}
\label{eqn:spectrum}
h_{n\ell} = h_H + \a\p{h_L + n} + \ell + O(1/c), \qquad
\bar{h}_{n\ell} = h_H + \a\p{h_L + n} + O(1/c),
\end{align}
where $h_H = \frac{\Delta_H}{2}$ and $h_L = \frac{\Delta_L}{2}$. They are precisely the modes $\phi_{n\ell}(t,\th) = \mathcal{C}_{n\ell} e^{-i\w_{n\ell} t} e^{i\ell\th}$ from \S\ref{sec:scalar-propagator}, and one can check that their energies $\Delta_{n\ell} = h_{n\ell} + \bar{h}_{n\ell}$ are exactly $\Delta_H$ plus the excitation energies $\w_{n\ell}$ that we found in (\ref{eqn:quantization}). Moreover, the normalization coefficients $\mathcal{C}_{n\ell}$ given by (\ref{eqn:fourier-coefficients}) are OPE coefficients: they are the amplitudes to produce states with $n$ and $\ell$ corresponding to a Fourier mode of $\phi$ on the boundary. 

The emergent picture is that the mode sum $G_N(t,\th) \sim \sum_{n\ell} \mathcal{C}_{n\ell}^2 e^{-i\w_{n\ell} t} e^{i\ell\th}$ in (\ref{eqn:wightman2}) seems to be an s-channel OPE that includes only the contributions of the primaries $\mathcal{O}_{n\ell}$. In what follows, we will substantiate this interpretation by computing the correlator (\ref{eqn:correlator-def}) in both the s and t channels, at leading order in large $c$. The former will reproduce the sum over modes, while the latter will make contact with the method of images.

\subsection{The s Channel: Double-Twist Bonanza}

We begin by inserting in (\ref{eqn:correlator-def}) a resolution of the identity. Because the CFT Hilbert space $\mathcal{H}_{\mathrm{CFT}}$ decomposes as a direct sum of irreducible representations of (two copies of) the Virasoro algebra, the identity can be written as a sum of the orthogonal projections $\mathcal{P}_{h, \bar{h}}$ onto each of these representations. Each one is spanned by a primary state $\ket{h, \bar{h}}$ and all of its descendants, and factorizes into holomorphic and anti-holomorphic pieces:
\begin{align}
\label{eqn:resolution}
\mathcal{H}_{\mathrm{CFT}} = \bigoplus_{h, \bar{h}} \Big( M(c, h) \otimes M(c, \bar{h}) \Big), \qquad 
\one = \sum_{h, \bar{h}} \p{\mathcal{P}_{h} \otimes \bar{\mathcal{P}}_{\bar{h}}}.
\end{align}
As discussed in \cite{Fitzpatrick:2014vua}, at large $c$ each irrep $M(c, h)$ admits an approximately orthogonal ``semiclassical graviton basis'' comprised of strings of Virasoro generators $L_{-m}$ ($m \geq 1$) acting on $\ket{h}$ to excite descendants (or ``boundary gravitons''). The projection is
\begin{align}
\label{eqn:projector}
\mathcal{P}_{h} &= \sum_{\{m_i, k_i\}}
\frac{L_{-m_1}^{k_1} \cdots L_{-m_r}^{k_r} \ket{h}\!\bra{h} L_{m_r}^{k_r} \cdots L_{m_1}^{k_1}}{\mel{h}{L_{m_r}^{k_r} \cdots L_{m_1}^{k_1} L_{-m_1}^{k_1} \cdots L_{-m_r}^{k_r}}{h}}, \qquad m_i, k_i \in \N,
\end{align}
where without loss of generality we take $m_1 > \cdots > m_r \geq 1$.

Resolving the identity between the two light fields in (\ref{eqn:correlator-def}) and using (\ref{eqn:projector}), we find 
\begin{align}
\label{eqn:s-channel-OPE}
G_{N}(z,\bar{z}) &= \sum_{h, \bar{h}} \frac{\big\langle \mathcal{O}_H(\infty) \mathcal{O}_L(1) \p{\mathcal{P}_h \otimes \bar{\mathcal{P}}_{\bar{h}}} \mathcal{O}_L(z, \bar{z}) \mathcal{O}_H(0) \big\rangle}{\big\langle \mathcal{O}_H(\infty) \mathcal{O}_H(0) \big\rangle} = 
\sum_{h, \bar{h}} \mathcal{V}_{h}(z) \bar{\mathcal{V}}_{\bar{h}}(\bar{z}),
\end{align}
where the semiclassical Virasoro conformal blocks $\mathcal{V}_h(z)$ are given by
\begin{align}
\label{eqn:virasoro-blocks}
\mathcal{V}_h(z) &= \sum_{\{m_i, k_i\}} \frac{\bra*{\mathcal{O}_H(\infty) \mathcal{O}_L(1) L_{-m_1}^{k_1} \cdots L_{-m_r}^{k_r}}\ket*{h}\!\bra*{h}\ket*{L_{m_r}^{k_r} \cdots L_{m_1}^{k_1} \mathcal{O}_L(z) \mathcal{O}_H(0)}}{\mel{h}{L_{m_r}^{k_r} \cdots L_{m_1}^{k_1} L_{-m_1}^{k_1} \cdots L_{-m_r}^{k_r}}{h} \big\langle \mathcal{O}_H(\infty) \mathcal{O}_H(0) \big\rangle}.
\end{align}
The sum runs unrestricted over the $m_i$ and $k_i$, but it may be organized into sectors of fixed ``graviton number'' $k = \sum_i k_i$. Accordingly, within each sector there is a $k$-graviton contribution $\mathcal{V}_h^{(k)}(z)$ to the full block, which is then given by $\mathcal{V}_h(z) = \sum_{k=0}^{\infty} \mathcal{V}_h^{(k)}(z)$.

Let us begin with the zero-graviton sector, corresponding to primary exchange. The block (\ref{eqn:virasoro-blocks}) reduces to three-point functions of the form $\langle \mathcal{O}_H \mathcal{O}_L \mathcal{O}_h \rangle$. Since their functional form is fixed by conformal symmetry, we can determine $\mathcal{V}_h^{(0)}(z)$ up to a constant:
\begin{align}
\label{eqn:0-graviton}
\mathcal{V}_h^{(0)}(z) = \frac{\big\langle \mathcal{O}_H(\infty) \mathcal{O}_L(1) \mathcal{O}_h(0)\big\rangle \big\langle\mathcal{O}_h(\infty) \mathcal{O}_L(z) \mathcal{O}_H(0) \big\rangle}{\big\langle\mathcal{O}_h(\infty) \mathcal{O}_h(0) \big\rangle \big\langle\mathcal{O}_H(\infty) \mathcal{O}_H(0)\big\rangle} \propto 
z^{h - h_H - h_L}.
\end{align}

To handle $\mathcal{V}_h^{(k)}(z)$ for $k > 0$, one uses the Virasoro algebra and the commutator
\begin{align}
\label{eqn:primary-commutator}
[L_m, \mathcal{O}_h(z)] = h(1+m) z^m \mathcal{O}_h(z) + z^{1+m} \partial_z \mathcal{O}_h(z)
\end{align}
to move the Virasoro generators past one another and the other primaries in (\ref{eqn:virasoro-blocks}) until they annihilate the vacuum. Due to (\ref{eqn:primary-commutator}), the numerator of $\mathcal{V}_h^{(k)}(z)$ will apply a differential operator to $\mathcal{V}_h^{(0)}(z)$ that produces a polynomial in $h - h_H - h_L$. But the present OPE is dominated by the Virasoro blocks of the $\mathcal{O}_{n\ell}$, and from (\ref{eqn:spectrum}) we have $h - h_H - h_L = O(c^0)$; hence the numerator of $\mathcal{V}_h^{(k)}(z)$ will be independent of $c$. Meanwhile, the normalizing denominator will be a degree-$k$ polynomial in $h$ and $c$; therefore overall we have $\mathcal{V}_h^{(k)}(z) = O(c^{-k})$. Thus at large $c$, the only unsuppressed contribution to the Virasoro block of a heavy-light double-twist primary comes from the exchange of the primary itself.

It is interesting to note that the same insight can be gleaned from the global conformal blocks of $\mathcal{O}_h$. These were worked out by Dolan and Osborn \cite{Dolan:2000ut}, and take the form
\begin{align}
\mathcal{F}_h(h_H, h_L; z) = z^{h - h_H - h_L} {}_2 F_1 \big[h - h_H + h_L, h - h_H + h_L; 2h; z \big].
\end{align}
The first two arguments of the ${}_2 F_1$ are $O(c^0)$, while the third, $2h_{n\ell} = \Delta_H + O(c^0)$, scales with $c$. By studying the asymptotics of ${}_2 F_1 \big[a, b; c + \l; z \big]$ as $\l \too \infty$ with $a, b, c, z$ fixed \cite{Watson:1918abc}, one finds that the hypergeometric part of the block is asymptotic to 1 at large $c$.

In any case, the double-twist Virasoro blocks reduce to ``scaling blocks:''
\begin{align}
\label{eqn:scaling-blocks}
\mathcal{V}_h(z) \bar{\mathcal{V}}_{\bar{h}}(\bar{z}) \approx \mathcal{V}_h^{(0)}(z) \bar{\mathcal{V}}_h^{(0)}(\bar{z}) = C_{HLh}^2 z^h \bar{z}^{\bar{h}} \abs{z}^{-\Delta_H - \Delta_L}.
\end{align}
The constants $C_{HLh}$ are precisely the OPE coefficients $\mathcal{C}_{n\ell}$ that we have worked out. Their knowledge allows us to complete the calculation: taking $h = h_{n\ell}$ as in (\ref{eqn:spectrum}), we substitute the OPE data (\ref{eqn:fourier-coefficients}) and the Virasoro blocks (\ref{eqn:scaling-blocks}) into the OPE (\ref{eqn:s-channel-OPE}). The result is
\begin{equation}
\begin{aligned}
\label{eqn:correlator-s1}
G_N(z, \bar{z}) &= \sum_{\ell \in \Z} \sum_{n \in \N} \mathcal{C}_{n\ell}^2 \abs{z}^{\p{\a - 1}\Delta_L + 2\a n} z^{\ell} = \frac{\a^{2\Delta_L - 1}}{2\pi \G(\Delta_L)} \sum_{\ell \in \Z} s_{\ell}(z,\bar{z}) \abs{z}^{\p{\a - 1} \Delta_L} z^{\ell}, \\
s_{\ell}(z, \bar{z}) &= \p{\frac{\G(\Delta_L + |\ell|/\a)}{\G(1 + |\ell|/\a)}} {}_2 F_1 \big[\Delta_L, \Delta_L + |\ell|/\a; 1 + |\ell|/\a; \abs{z}^{2\a} \big].
\end{aligned}
\end{equation}
This exactly matches the bulk expression (\ref{eqn:wightman2}), as well as (\ref{eqn:main-result}) on the unit circle.

Another way to think about this is that the field
$\phi$ is quasi-free and of low dimension in the vacuum. One would expect the OPE to be dominated by primaries of the form ${\mathcal O}_H\partial^{[k]}{\mathcal O}_L$ with no derivatives on the heavy operator \cite{Berenstein:2021pxv}. The extra terms in the $1/c$ expansion are due to recoil, but recoil is suppressed when one has a heavy operator.

\subsection{The t Channel: Vacuum Dominance}


The calculation in the t channel turns out to be both simple and conceptually intriguing. Here the OPE is dominated by the Virasoro vacuum block, which includes the stress tensor and captures the exchange of boundary gravitons between the defect and the scalar.

The semiclassical Virasoro vacuum block was worked out by Fitzpatrick, Kaplan, and Walters \cite{Fitzpatrick:2014vua, Fitzpatrick:2015zha}, who showed that the t-channel Virasoro blocks for a heavy-light correlator essentially reduce to global conformal blocks, but with the light fields evaluated in a new set of coordinates, $w(z) = z^{\a}$, so that $\mathcal{O}_L(z) \too \p{w'(z)}^{h_L} \mathcal{O}_L(w)$:
\begin{align}
\label{eqn:vacuum-block}
\mathcal{V}_h(z) = z^{\p{\a - 1} h_L} \p{\frac{1 - z^{\a}}{\a}}^{h - 2h_L} {}_2 F_1 \big[h, h; 2h; 1 - z^{\a}].
\end{align}
The basic idea is that the heavy operators in the correlator produce a new, ``thermal'' background geometry for the light operators \cite{Fitzpatrick:2015zha}, in the same way that a massive particle or a BTZ black hole causes the bulk geometry to backreact and modify the motion of a light probe in its presence. It is remarkable that some notion of thermality is reflected in the Virasoro blocks themselves, even below the BTZ threshold. (We will explore some other features of the scalar propagator that hint at thermal behavior in \S\ref{sec:thermal-physics} below.)



In any case, we take $h = \bar{h} = 0$ in (\ref{eqn:vacuum-block}) to obtain the vacuum block $\mathcal{V}_0(z)$. That this block dominates the OPE means that we approximate $G_N(z, \bar{z})$ by $\mathcal{V}_0(z) \bar{\mathcal{V}}_0(\bar{z})$:
\begin{align}
\label{eqn:correlator-t1}
\mathcal{V}_0(z) = z^{\p{\a - 1} h_L} \p{\frac{1 - z^{\a}}{\a}}^{-2h_L} \implies
G_N(z, \bar{z}) \approx \abs{z}^{\p{\a - 1} \Delta_L} \abs{\frac{1 - z^{\a}}{\a}}^{-2\Delta_L}.
\end{align}
Passing from $(z,\bar{z})$ to $(t,\th)$ coordinates and restricting to the region $|z-1| < 1$, we find
\begin{align}
\label{eqn:correlator-t2}
G_N(t,\th) \approx \bigg[\frac{2}{\a^2} \Big(\! \cos\p{\a t} - \cos\p{\a \th}\Big)\bigg]^{-\Delta_L}, \qquad
G_N(0, \th) \approx \bigg[\frac{2}{\a} \sin\p{\frac{\a\th}{2}} \bigg]^{-2\Delta_L}.
\end{align}

\begin{figure}[t]
\centering
\[
\begin{gathered}
\begin{tikzpicture}
    \begin{feynman}
    \vertex (a);
    \vertex [left = 0.1cm of a] (a1) {\( \Delta_L \)};
    \vertex [above left = 1.5cm and -0.1cm of a] (b) {\(\mathcal{O}_L(z, \bar{z})\)};
    \vertex [below  left = 1.5cm and -0cm of a] (d) {\(\mathcal{O}_L(1)\)};
    \vertex [right = 3cm of a] (f);
    \vertex [right = 0.1cm of f] (g) {\( \Delta_H \)};
    \vertex [above right = 1.5cm and -0cm of f] (c) {\(\mathcal{O}_H(\infty)\)};
    \vertex [below right = 1.5cm and -0cm of f] (e) {\(\mathcal{O}_H(0)\)};
    \diagram* {
        (b) -- (a) -- (d),
        (a) -- [scalar, edge label = {$\mathcal{V}_{\one}(z, \bar{z})$}, 
        edge label'=\(2/c\)
        ] (f),
        (c) -- [plain] (f) -- [plain] (e),
    };
    \end{feynman}
\end{tikzpicture}
\end{gathered}
\scalebox{1.6}{$\displaystyle \;\;\; = \;\;\; \sum_{n\ell}\!\!$}
\begin{gathered}
\begin{tikzpicture}
    \begin{feynman}
    \normalsize
    \vertex (a);
    \vertex [above left = 0.1cm and 1.5cm of a] (a1) {\( \mathcal{O}_L(z, \bar{z}) \)};
    \vertex [above right = 0.1cm and 1.5cm of a] (a2) {\( \mathcal{O}_H(\infty) \)};
    \vertex [above = 0.1cm of a] (c1) {$\mathcal{C}_{n\ell}$};
    \vertex [below = 3cm of a] (b);
    \vertex [below left = 0.1cm and 1.5cm of b] (b1) {\( \mathcal{O}_L(1) \)};
    \vertex [below right = 0.1cm and 1.5cm of b] (b2) {\( \mathcal{O}_H(0) \)};
    \vertex [below = 0.1cm of b] (c2){$\mathcal{C}_{n\ell}$};
    \diagram* {
    (a1) -- (a) -- (a2),
    (a) -- [scalar, edge label = $\;\mathcal{O}_{n\ell}$] (b),
    (b1) -- (b) -- (b2),
    };
    \end{feynman}
\end{tikzpicture}
\end{gathered}
\]
\vspace{-11pt}
\caption{These diagrams illustrate the crossing equation and the kinematics of our OPEs. The t channel (left) is dominated by the vacuum block, while the s channel (right) is dominated by the double-twist primaries. The $\frac{2}{c}$ on the left is a normalization: in the OPE $T(z)T(z') \sim \frac{c/2}{(z-z')^4} + \cdots$, the coefficient of the leading singularity is the norm-squared of $T$ in the Zamolodchikov metric \cite{Zamolodchikov:1986gt}. On the right, we have labeled the OPE coefficients that correspond to the 3-point functions $\big\langle \mathcal{O}_H \mathcal{O}_L \mathcal{O}_{n\ell} \big\rangle$. Both sides can also be interpreted as bulk Witten diagrams \cite{Hijano:2015qja}.}
\label{fig:feynman-diagram}
\end{figure}
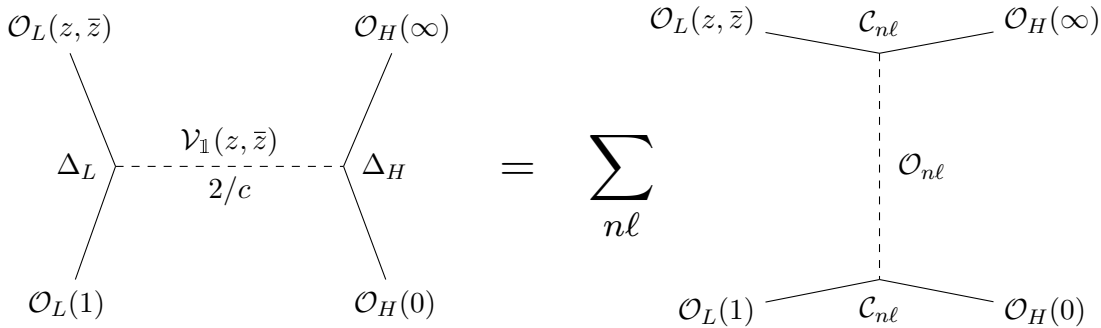

Several comments on these results are now in order.
\begin{enumerate}
\item The s- and t-channel results (\ref{eqn:correlator-s1}) and (\ref{eqn:correlator-t1}) have identical singular expansions in $|1-z|$, and both are consistent with (\ref{eqn:bulk-expansion}). This verifies that they are related by crossing symmetry, and gives a nontrivial check of the crossing equation:
\begin{align}
\frac{1}{\a \G(\Delta_L)^2} \sum_{\ell \in \Z} \sum_{n \in \N} 
\frac{\G(\Delta_L + n) \G(\Delta_L + n + |\ell|/\a)}{\G(1+n) \G(1 + n + |\ell|/\a)} \abs{z}^{2\a n} z^{\ell} \sim \abs{1 - z^{\a}}^{-2\Delta_L}.
\end{align}
(See Fig. \ref{fig:feynman-diagram}.) The main lesson here is that the asymptotic behavior of a complicated sum on the LHS, which arises naturally in the bulk, controls a singularity whose form is manifest on the RHS, and which is more natural on the boundary.
\item The result (\ref{eqn:correlator-t2}) reproduces the dominant term in the $\mathrm{AdS}_3/\Z_N$ propagator (\ref{eqn:images-result}--\ref{eqn:images-time}) obtained by the method of images; consequently it also agrees with the geodesic approximation, up to a normalization constant: $\mathcal{V}_0(z) \sim e^{-m \mathcal{L}[\g_0]}$. This highlights a deep connection between Virasoro blocks and bulk geodesics explored in \cite{Asplund:2014coa, Hijano:2015qja}. There it is explained how the vacuum block arises from bulk Witten diagrams, and how the geodesic length it computes arises from a certain worldline action.
\item The factors of $\a$ in (\ref{eqn:correlator-t2}), relative to the case of pure $\mathrm{AdS}_3$, are manifestations of the ``thermal'' coordinates of \cite{Fitzpatrick:2015zha}. This is the boundary analog of the rescaling between covering and quotient coordinates in $\mathrm{AdS}_3/\Z_N$ that we discussed in \S\ref{sec:method-of-images}.
\item The vacuum block is a subtle object. While we consider here the limit $c \too \infty$ with $h_L/c \too 0$ and $h_H/c$ fixed, there are also other regimes; for instance, $c \too \infty$ with $(h_L h_H)/c$ fixed but $h_L/c \too 0$ and $h_H/c \too 0$. In that limit, considered in \cite{Fitzpatrick:2014vua}, the vacuum block exponentiates the global block of the stress tensor:
\begin{align}
\mathcal{V}_0^{\mathrm{alt}}(z) = \exp\p{\frac{2h_L h_H}{c} z^2 {}_2 F_1[2, 2; 4; z]}.
\end{align}
Here, unlike in (\ref{eqn:correlator-t1}), the fact that the block exponentiates is manifest. But in a power expansion in $|1-z|$, only the leading (universal) and first subleading terms of $\mathcal{V}_0(z)$ and $\mathcal{V}_0^{\mathrm{alt}}(z)$ agree. 
See Fig. \ref{fig:feynman-diagram}, and compare Fig. 7 of \cite{Fitzpatrick:2015zha}.
\end{enumerate}

\section{The Onset of Thermal Physics}
\label{sec:thermal-physics}

Physics in conical $\mathrm{AdS}_3$ is not thermal. There is no horizon, no black hole, no Hawking temperature, and no entropy; moreover, the dual CFT state $\ket{\mathcal{O}_H}$ is pure. Nevertheless, we shall show that physics in conical $\mathrm{AdS}_3$ has certain thermal features. In particular, we will argue that $G_N(\th)$ decays as $N$ increases, implying that correlations become suppressed as we increase the defect's mass and approach the BTZ threshold. Just as thermal correlation lengths get shorter as one increases the temperature (i.e. the energy density), correlations in conical $\mathrm{AdS}_3$ at fixed length decrease as we increase the energy of the defect state.

\subsection{In Search of Lost Temperature} 

\begin{figure}[t]
    \centering
    \includegraphics[width=.9\linewidth, trim={35 0 5 0}, clip]{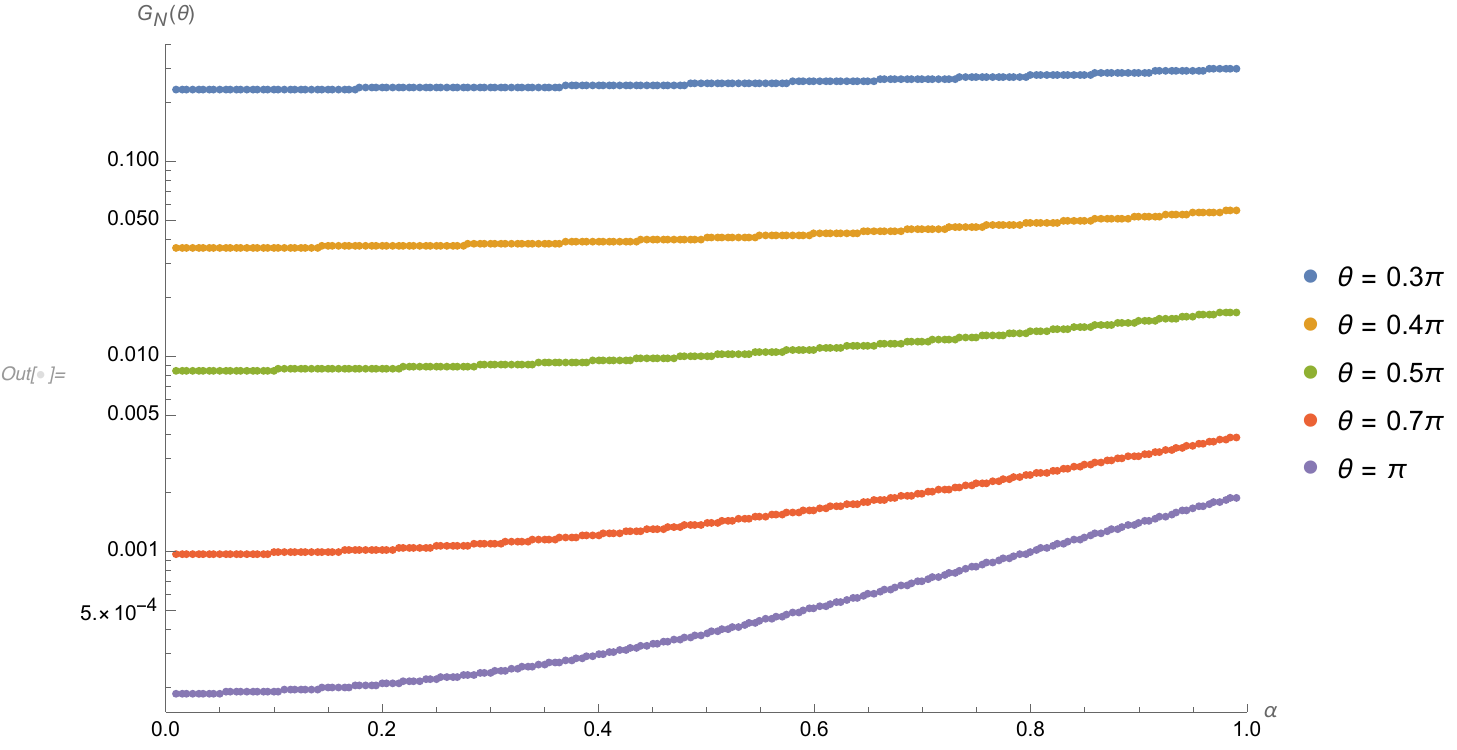}
    \caption{The propagator $G_N(\th)$ is monotonically increasing in $\a = \frac{1}{N}$ (hence decreasing in $N$) throughout the full range of defect masses $\a \in (0,1]$, for selected values of $\th$ and $\Delta = 3.25$.}
    \label{fig:decay-n}
\end{figure}

We have confirmed numerically that $G_N(\th)$ decreases monotonically with $N$. This behavior can already be seen in Fig. \ref{fig:propagators}, which shows $G_N(\th)$ for several values of $N$. It is even more apparent in Fig. \ref{fig:decay-n}, which plots $G_N(\th)$ against $\a = \frac{1}{N}$ directly, for several values of $\th$. This decay of correlations describes a ``screening'' effect, suggestive of thermality, which we expect to persist even above the BTZ threshold. We can also see that $G_N(\th)$ is decreasing in $N$ from its bulk expansion (\ref{eqn:bulk-expansion}), whose leading $N$-dependent term scales like $N^{-2}$.

A more transparent argument starts from the t-channel result (\ref{eqn:correlator-t2}). We have
\begin{align}
G_N(\th) \approx \qty[2N \sin\p{\frac{\th}{2N}}]^{-2\Delta} \implies
\frac{\partial G_N}{\partial N} = -\frac{2\Delta}{N} G_N(\th) \qty[1 - \frac{\th}{2N} \cot\p{\frac{\th}{2N}}] \leq 0,
\end{align}
where the inequality follows because $G_N(\th)$ and the term in square brackets are both positive. Thus the dominant image is monotonically decreasing in $N$. And for sufficiently large $N$, the contributions of the other images are exponentially suppressed: therefore the full image sum is decreasing in $N$ as we approach the black hole threshold.

\subsection{The Response Function} 

So far, we have studied the system by asking how the background geometry affects the scalar. But the ground state of $\phi$ in this background is also affected if we perturb the Hamiltonian of the theory by adding at the boundary a term linear in $\phi$. We can use linear response theory to quantify the effect; this is another way to understand what the defect does to the scalar. To find the response, we pass to Euclidean signature in the dual CFT, add a constant source $J\int_0^{2\pi} \phi(\t,\th)\, \dd\theta$ to the Hamiltonian, and study the state of the scalar. The system's response to this deformation, measured by the expectation value of $\phi$, is obtained \cite{Kubo:abc} by integrating the boundary propagator:\footnote{To be more explicit, one should start in Lorentzian signature and integrate the response function $\chi(t_1,t_2,\th_1,\th_2) = -i \Theta(t_1-t_2) \big\langle [\phi(t_1,\th_1), \phi(t_2,\th_2)]\big\rangle_{N} = \Theta(t_1-t_2) G^+_N(t_1-t_2,\th_1 - \th_2)$ over the $\mathrm{AdS}_3$ boundary. One sets $t_1 = \th_1 = 0$, renames $t \equiv -t_2$ and $\th \equiv -\th_2$, and then rotates to Euclidean signature to find (\ref{eqn:response-setup}).}
\begin{align}
\label{eqn:response-setup}
\left< \frac{\d \phi}{\d J} \right>_N \equiv \frac{\d}{\d J} \Bigg|_{J = 0} \! \mel{\mathcal{O}_H}{\phi(\t = 0, \th = 0)}{\mathcal{O}_H} = \int_0^{\infty} \dd \t \int_0^{2\pi} \dd\th\, G_N^+(\t,\th).
\end{align}
The same analysis was done in conformal perturbation theory for the CFT vacuum in \cite{Berenstein:2014cia}. There the $\mathrm{AdS}$ calculation was also done---one finds the s-wave mode of $\phi$ by introducing a boundary source and using regularity at the origin---and it agrees with the CFT. 

To evaluate the integral (\ref{eqn:response-setup}), we substitute (\ref{eqn:wightman3}) directly: the integral over $\th$ kills all of the modes with $\ell \neq 0$ by symmetry, leaving only the $\ell=0$ mode. The angular integral is then trivial, and the integral over $\t$ is evaluated by setting $u = e^{-\t/N}$:
\begin{equation}
\label{eq:staticresponse}
\begin{aligned}
\left< \frac{\d \phi}{\d J} \right>_N &= N^{1-2\Delta} \int_0^{\infty} \dd \t\, e^{-\Delta \t/N} {}_2 F_1 \Big[ \Delta,\, \Delta;\, 1;\, e^{-\frac{2 \t}{N}} \Big] = \\
&= N^{2-2\Delta} \int_0^1 \dd u\, u^{\Delta - 1} {}_2 F_1 \big[ \Delta,\, \Delta;\, 1;\, u^2 \big] = \frac{N^{2-2\Delta}\G\p{\frac{1-\Delta}{2}} \G\p{\frac{\Delta}{2}}}{2^{1+2\Delta} \G\p{\frac{1+\Delta}{2}} \G\p{1 - \frac{\Delta}{2}}}.
\end{aligned}
\end{equation}
The integral above converges only in the strip $0 < \mathrm{Re}\{\Delta\} < 1$. But the result is analytic in $\Delta$, so it determines $\big\langle \phi \big\rangle_N$ uniquely for larger $\Delta$. (The gravity computation, as in \cite{Berenstein:2014cia}, automatically takes this continuation into account.) Note that $\big\langle \phi \big\rangle_N$ has resonances: the response vanishes when $\Delta$ is even, and diverges when $\Delta$ is odd. These singularities arise from logarithmic subdivergences in conformal perturbation theory and lead to secular divergences for certain time-dependent questions (see, for instance, \cite{Berenstein:2016avf}). But these details are immaterial when we compare (\ref{eq:staticresponse}) to the response in pure $\mathrm{AdS}_3$:
\begin{align}
\label{eqn:response-result}
\Phi_N \equiv \left< \frac{\d \phi}{\d J} \right>_N \Bigg/\left< \frac{\d \phi}{\d J} \right>_1 = N^{2-2\Delta} = \qty[\frac{12}{c} \big(M_* - M\big)]^{\Delta - 1}, \qquad M_* = \frac{c}{12}.
\end{align}

The factors of $N$ are easy to explain. For both pure $\mathrm{AdS}_3$ and in $\mathrm{AdS}_3/\Z_N$, the perturbation is in the s-wave of $\phi$: the only difference between the two is the relative rescaling of the boundary coordinates by $N$. This leads to the familiar factor of $N^{-2\Delta}$ (discussed also in \S\ref{sec:method-of-images} above) in $\Phi_N$. The rescaling of the integral measure in (\ref{eqn:response-setup}) between both coordinate systems provides an extra factor of $N^2$. The response for arbitrary $N \in \R_+$ is then just the na\"ive analytic continuation of the integer-$N$ case. We write $\Phi_N$ in terms of the difference $M_* - M$ to highlight the presence of critical exponents near $M_*$.

The time-dependent quantity $G_N^{(\mathrm{s})}(\t) = e^{-\Delta\t/N} N^{1-2\Delta} {}_2 F_1[\Delta,\,\Delta;\, 1;\, e^{-2\tau/N}]$ in the top line of \eqref{eq:staticresponse} is the Green's function for the s-wave modes. It shows that the OPE coefficients of the $\ell = 0$ excited states scale as $N^{1-2\Delta}$, and that the density of states increases by a factor of $N$ relative to pure $\mathrm{AdS}_3$. In Lorentzian time, $G_N^{(\mathrm{s})}(t)$ tells us that signals sent out from the boundary are delayed in their return by a factor of $N$ relative to $\mathrm{AdS}_3$. (See \cite{Berenstein:2020cll} for a related analysis in pure $\mathrm{AdS}$.) When $N$ becomes large and we are near the BTZ threshold, we might not see the signal come back before an experiment is carried out.

\subsection{A Curious Phase Transition} 

The power law behavior (\ref{eqn:response-result}) signals a nontrivial second-order phase transition in $G_N(\th)$, for which the response ratio $\Phi_N$ acts as a proxy. As long as $\Delta$ is not an integer, $\Phi_N$ fails to be analytic in $M$ near the BTZ threshold $M_*$. The critical exponent $\Delta - 1$ dictates whether $\Phi_N$ vanishes or diverges at $M_*$. For $\Delta = \Delta_+ > 1$, the response dies off as $N$ grows large, and in the massless BTZ geometry the scalar perturbation is not felt at all: there is no response. Above $M_*$, the perturbation becomes irrelevant in the black hole throat. But in the range $\Delta = \Delta_- \in (0,1)$, corresponding to a scalar with $-1 < m^2 < 0$ quantized using modified boundary conditions, $\Phi_N$ diverges as $N \too \infty$: the scalar is relevant in the black hole throat. Thus the critical-mass defect becomes sensitive to scalar deformations of the boundary only when the latter constitute relevant perturbations, and even then only for the choice $\Delta = \Delta_-$ of alternate boundary conditions.

Critical behavior can also be seen in the OPE coefficients $\mathcal{C}_{n\ell}$ near the transition, as these also scale with powers of $N$. Moreover, the spectrum (\ref{eqn:spectrum}) shows that the gap between excited states closes as $\frac{1}{N}$, so small fluctuations in energy in conical $\mathrm{AdS}_3$ can have much more entropy than they can near the $\mathrm{AdS}$ vacuum. We expect that these fluctuations eventually become heavy enough that they form a horizon. The gap between modes of different $\ell$ is still of order $1$, rather than $\frac{1}{N}$, as can be seen from (\ref{eqn:spectrum}). Not much entropy can be stored in these states, even though they individually become degenerate. To each such angular momentum mode, we associate a tower of states with energies $\frac{1}{N} \p{\Delta + 2n}$: this is the density of states of a single chiral boson on a circle of radius $N$. The specific heat of these states scales like $N$ times the number of light species; in this sense, the specific heat near the transition is also divergent and typical of a continuous phase transition.

\section{Applications to Entanglement Entropy}
\label{sec:applications}

As an application, we can use our propagator to approximate the entanglement entropy of a boundary interval in conical $\mathrm{AdS}_3$. We begin by arguing that in certain regimes, the twist fields arising in the 2D CFT replica trick should not be too heavy. This proposal simplifies several calculations and reproduces several known results, and we show that it also agrees with the Ryu--Takayanagi (RT) formula in $\mathrm{AdS}_3/\Z_N$ in the large-$c$ limit.

\subsection{Entropy from Light Twist Fields}
\label{sec:entropy-background}

Consider a holographic 2D CFT in its vacuum state $\rho = \ket{0} \! \bra{0}$, and let $A = [u,v]$ be a spatial interval of length $\ell$. The reduced density matrix $\rho_A$ represents the state seen by an observer with access only to $A$, and the entanglement entropy $S_A[\rho] \equiv -\Tr\qty[\rho_A \ln \rho_A]$ measures the entanglement between degrees of freedom in $A$ and those in the rest of the system. One computes $S_A[\rho]$ by considering the R\'enyi entropies $S_A^{(n)}[\rho]$, a family of related quantities labeled by the R\'enyi index $n \in \R_+$ and defined by
\begin{align}
S_A^{(n)}[\rho] = \frac{1}{1-n} \ln \Big( \Tr \big[\rho_A^{n} \big] \Big), \qquad S_A[\rho] = \lim_{n \to 1} S_A^{(n)}[\rho] = -\frac{\partial}{\partial n} \Big( \Tr \big[\rho_A^{n} \big] \Big) \eval_{n = 1}.
\end{align}
Calabrese and Cardy \cite{Calabrese:2009qy} used the replica trick to express $\Tr[\rho_A^n]$ as a two-point function of scalar primary operators $\s_{n},\, \bar{\s}_{n}$ called twist fields, with dimensions $\Delta_{n} = \frac{c}{12} \p{n - \frac{1}{n}}$. Conformal invariance then determines the entropy $S_A[\rho]$ in the CFT vacuum:
\begin{align}
\label{eqn:entropy-poincare}
\Tr[\rho_A^{n}] = \big\langle \s_{n}(u) \bar{\s}_{n}(v) \big\rangle = \frac{1}{|u - v|^{2\Delta_{n}}} = \ell^{-\frac{c}{6} \p{n - \frac{1}{n}}} \implies S_A[\rho] = \frac{c}{3} \ln(\ell).
\end{align}

In an excited state like $\rho = \ket{\mathcal{O}_H}\bra{\mathcal{O}_H}$, the calculation is much harder because we need to evaluate a four-point function of heavy operators: $\mathrm{Tr}[\rho_A^n] \propto \big\langle \mathcal{O}_H \s_n \bar{\s}_n \mathcal{O}_H \big\rangle$. But on comparing the dimensions $\Delta_H = \frac{c}{12}\p{1 - \frac{1}{N^2}}$ and $\Delta_{n} = \frac{c}{12} \p{n - \frac{1}{n}}$, we see that in the limit $n \too 1$ the twist fields become lighter while the defect remains heavy \cite{Banerjee:2016qca}.

We are led to believe that a calculation where the twist fields are light might be dominated by the propagator we have computed. This is because we neglect backreaction, so the correlator will behave much like the two-point function of $\phi$. (Notice that this is a CFT idea, rather than a gravitational path integral.) When the twist fields are close to each other, we can consider their OPE; there we expect that the identity dominates. What about fields like $\phi$? Its ground state in the defect background is characterized by a vanishing expectation value: $\big\langle \phi \big\rangle_{N} \propto \big\langle \mathcal{O}_H \mathcal{O}_L \mathcal{O}_H \big\rangle = 0$. The vanishing of this three-point function tells us that the expansion $\mathcal{O}_L(z) \mathcal{O}_H(0) \sim z^{-\Delta} C_{HLH} \mathcal{O}_H(0) + \cdots$ has vanishing OPE coefficient $C_{HLH} = 0$. So for the scalar, too, only the vacuum block contributes.

The upshot is that, upon taking the twist fields to be light,\footnote{We are taking $\sigma_n$ and $\bar{\sigma}_n$ to be quasi-free fields, which makes them dual to free scalars in the bulk \cite{Duetsch:2002hc}.} their correlator in the defect background becomes our propagator: $\mathrm{Tr}[\rho_A^n] = G_N(\th)$. Thus the R\'enyi entropies $S_N^{(n)}(\th)$ for an interval of opening angle $\th$ in conical $\mathrm{AdS}_3$ are well approximated by
\begin{align}
\label{eqn:quasifree-entropy}
S_N^{(n)}(\th) = \frac{\ln \big\langle \s_{n}(0,\th)\, \bar{\s}_{n}(0,0) \big\rangle_{N}}{1-n} = \frac{\ln \big( G_N(\th) \big)}{1-n}, \qquad \Delta \too \Delta_{n} = \frac{c}{12} \p{n - \frac{1}{n}}.
\end{align}

\subsection{A Few Consistency Checks}
\label{sec:consistency-checks}

As a first test of our proposal, we consider short distances. At small opening angles, the universal divergence (\ref{eqn:mBTZ}) of $G_N(\th)$ becomes a logarithmic divergence in the entropy:
\begin{align}
\label{eqn:entropy-UV}
\Tr[\rho_A^{n}] = G_N(\th) \approx \frac{\th^{-2\Delta_{n}}}{2\pi} \implies S_N(\th) \approx \frac{c}{3} \ln(\th),
\end{align}
matching (\ref{eqn:entropy-poincare}). (In taking the limit $n \too 1$, we discarded an infinite constant associated with the normalization of $G_N(\th)$, to which entropies are not sensitive.) Next, consider the case of pure $\mathrm{AdS}_3$ in global coordinates. The propagator is given by (\ref{eqn:n=1}), so
\begin{align}
\label{eqn:entropy-n=1}
\Tr[\rho_A^{n}] = G_1(\th) = \frac{1}{2\pi} \qty[2 \sin\p{\frac{\th}{2}}]^{-2\Delta_{n}} \implies S_1(\th) = \frac{c}{3} \ln\qty[2 \sin\p{\frac{\th}{2}}].
\end{align}
This is once again the standard result. Both (\ref{eqn:entropy-UV}) and (\ref{eqn:entropy-n=1}) agree with the RT formula, but we will see below that this is far from trivial in more general situations.

To move beyond the universal features of the entropy, we can leverage the semiclassical vacuum block, which we considered in \S\ref{sec:CFT-interpretation}. From (\ref{eqn:correlator-t2}) we find
\begin{equation}
\label{eqn:entropy-semiclassics}
\begin{aligned}
\Tr[\rho_A^{n}] &= G_N(\th) \approx \frac{1}{2\pi} \qty[2 N \sin\p{\frac{\th}{2N}}]^{-2\Delta_{n}} \implies S_N(\th) \approx \frac{c}{3} \ln \qty[2N \sin\p{\frac{\th}{2N}}].
\end{aligned}
\end{equation}
This approximation, which is well controlled as long as $\th$ is small, shows explicitly how the entropy rises as the defect becomes more and more massive. 

When $N$ is close to enough to 1 that the defect mass is small (more precisely, we take $\eta \equiv 1 - \frac{1}{N^2} = \frac{12M}{c} \ll 1$), the difference in entropies between conical and pure $\mathrm{AdS}_3$ is
\begin{equation}
\begin{aligned}
\d S(\th) &\equiv S_N(\th) - S_1(\th) = \frac{c}{3} \ln \qty[N \frac{\sin(\th/2N)}{\sin(\th/2)}] \approx 2M \qty[1 - \frac{\th}{2} \cot\p{\frac{\th}{2}}].
\end{aligned}
\end{equation}
In terms of the geodesic length $\mathcal{L}(\th) \equiv 2 \ln \qty[2 \sin\p{\frac{\th}{2}}]$ in pure $\mathrm{AdS}_3$, we find that
\begin{equation}
\begin{aligned}
S_N(\th) = S_1(\th) + \delta S(\th) &\approx
\frac{c}{3} \ln\qty[2 \sin\p{\frac{\th}{2}}] + \eta \frac{c}{6} \qty[1 - \frac{\th}{2} \cot\p{\frac{\th}{2}}] = \\ &=
\frac{1}{4G} \qty[\L(\th) - \frac{\eta\th}{2} \L'(\th) + \eta] \approx \frac{1}{4G} \qty[\L\p{\th - \frac{\eta}{2}\th} + \eta].
\end{aligned}
\end{equation}
Thus for small defects, the entropy may be determined using the RT formula in pure $\mathrm{AdS}_3$. To account for the defect, one needs only to shift the RT surface so that it subtends the smaller angle $\th \too \th\p{1 - \frac{\eta}{2}}$, and then to add $\eta$ to the length.

\subsection{The RT Formula for Quotients}
\label{sec:RT}

The RT formula \cite{Ryu:2006ef} says that the entanglement entropy of a spatial boundary interval $A$ of a holographic 2D CFT in state $\rho$ is given by the length of the shortest spatial bulk geodesic $\g$ anchored to the endpoints of $A$ and homologous to $A$. In the form
\begin{align}
\label{eqn:RT-formula}
S_A[\rho] = \min_{\partial \g = \partial A} \frac{\L[\g]}{4G} = -\ln \p{e^{-\L[\g_{\mathrm{min}}]/4G}},
\end{align}
the RT formula reminds us of the formal similarity, observed also in \cite{Fischetti:2014zja}, between the entropy and a saddle-point path integral in the geodesic approximation.

In $\mathrm{AdS}_3/\Z_N$, all but two of the geodesics $\g_0, ..., \g_{N-1}$ between two boundary points are ``long'' and wind around the singularity. As shown in Fig. \ref{fig:winding}, only $\g_0$ and $\g_{N-1}$ are ever minimal---the former for smaller intervals ($\th < \pi$), the latter for larger ones ($\th \geq \pi$). We can find the entropy of an interval in $\mathrm{AdS}_3/\Z_N$ by substituting (\ref{eqn:geodesic-length}) into (\ref{eqn:RT-formula}):
\begin{equation}
\label{eqn:RT-entropy}
\begin{aligned}
S_N^{\mathrm{RT}}(\th) &= \frac{c}{3} \min \qty{ \ln\qty[2N \sin\p{\frac{\th}{2N}}],\, \ln\qty[2N \sin\p{\frac{2\pi - \th}{2N}}] } = \\ &= \begin{cases}
\frac{c}{3} \ln\qty[2N \sin\p{\frac{\th}{2N}}], & \th < \pi, \\
\frac{c}{3} \ln \qty[2N \sin\p{\frac{2\pi - \th}{2N}}], & \th \geq \pi \end{cases} \qquad (N \in \Z_+),
\end{aligned}
\end{equation}
At $\th = \pi$, the two geodesics exchange minimality, leading to a sharp corner in $S_N^{\mathrm{RT}}(\th)$. This trade-off, known as the RT phase transition, is a generic feature of setups with multiple competing bulk geodesics. It is now understood that the RT formula receives sizable corrections near the phase transition at $O\p{1/\sqrt{G}} = O(\sqrt{c})$ that smooth out the corner \cite{Marolf:2020vsi, Dong:2020iod, Dong:2021oad}. However, such corrections do not show up in our computation. 

\subsection{The Order of Limits Matters}
\label{sec:order-of-limits}

The RT formula inhabits two extremes: it lives in the limit $n \too 1$, and also at large $c$. But these limits do not commute, so our prescription must be handled carefully. In $\mathrm{AdS}_3/\Z_N$, for example, we find $S_N^{(n)}(\th)$ by substituting the propagator (\ref{eqn:images-result}) into (\ref{eqn:quasifree-entropy}):
\begin{align}
\label{eqn:quasifree-renyi}
S_N^{(n)}(\th) = \frac{\ln\big( G_N(\th) \big)}{1-n} = \frac{1}{1-n} \ln \qty{\frac{1}{2\pi} \sum_{k=0}^{N-1} \qty[2N \sin\p{\frac{\th + 2\pi k}{2N}}]^{-\frac{c}{6} \p{n - \frac{1}{n}}}}.
\end{align}
Na\"ively, the von Neumann entropy can be obtained from (\ref{eqn:quasifree-renyi}) by taking the limit $n \too 1$. If we did this, we would find a disastrous disagreement with the RT formula:
\begin{align}
\label{eqn:wrong-entropy}
S_N^{\mathrm{W}}(\th) = \frac{c}{3N} \sum_{k=0}^{N-1} \ln \qty[2N \sin\p{\frac{\th + 2\pi k}{2N}}] = \frac{1}{N} \sum_{k=0}^{N-1} \p{\frac{\L[\g_k(\th)]}{4G}} \neq S_N^{\mathrm{RT}}(\th).
\end{align}

Evidently $S_N^{\mathrm{W}}(\th)$ computes the average---not the minimal---geodesic length in $\mathrm{AdS}_3/\Z_N$. What went wrong is that while all $N$ geodesics contribute to the propagator, the contributions of the non-minimal ones should be strongly suppressed in the entropy at large $c$. According to the Lewkowicz--Maldacena construction \cite{Lewkowycz:2013nqa}, one should take the limit $c \too \infty$ before sending the replica number $n$ to 1, so that the only meaningful contribution to the sum in (\ref{eqn:quasifree-renyi}) comes from either $\g_0$ (for $\th < \pi$) or $\g_{N-1}$ (for $\th \geq \pi)$. If we neglect the other terms before sending $n \too 1$, we recover the RT result (\ref{eqn:RT-entropy}).

The order-of-limits issue discussed above is a classic one \cite{Fischetti:2014zja, Headrick:abc}. If we treat the limits correctly, we find a simple picture of holographic entanglement entropy: in certain regimes, it is just the logarithm of a scalar propagator. To be sure, our prescription has only a limited range of validity; nevertheless, it serves as a useful illustration of several related techniques for computing and understanding holographic entanglement.


\section{Discussion and Conclusions}
\label{sec:conclusions}

In this paper, we have presented a detailed study of the free scalar propagator in conical $\mathrm{AdS}_3$. We derived the propagator directly from the bulk equations of motion, explored its short-distance behavior, treated it by the method of images in $\mathrm{AdS}_3/\Z_N$, used CFT techniques to stress-test our results, diagnosed the system's approach to thermality, and considered an application to holographic entanglement entropy using twist fields.

There are still aspects of our work that we would like to understand better. One such point concerns the Virasoro vacuum block's lack of periodicity, which indicates that there are important states in the t-channel OPE that we have not considered. For instance, there are light multiparticle states formed solely from the scalar. It would be interesting to work out the light-light double-twist OPE coefficients, and to match the regular terms in the bulk expansion of the correlator arising from subdominant images. A similar discussion applies to twist fields: by letting them exchange states other than the identity, we might be able to resolve the order-of-limits ambiguity in the entropy. More generally, the corrections to this story at $O(1/c)$ should in principle be tractable, and may give some insight into thermality and the physics of boundary gravitons. In particular, one could try to capture the contributions of double-twist Virasoro descendants in the s-channel OPE: these arise from the Goldstone modes of the defect and can be interpreted as recoil effects \cite{Berenstein:2021pxv}.

The continuous phase transition experienced by $G_N(t,\th)$ near the BTZ threshold is one of its most intriguing features, and it would be interesting to understand it both above and below the transition. But the presence of a horizon in the BTZ geometry makes the analytic structure of one- and two-point functions there more subtle: it would be good to understand them in more detail. Similar questions can also be asked in higher-dimensional setups: see \cite{Berenstein:2022nlj}, for example, for some of the answers.

The average geodesic length obtained in (\ref{eqn:wrong-entropy}) also raises questions. While it is certainly not an entropy, it is related to a similar quantity called entwinement \cite{Balasubramanian:2014sra}. Both entwinement and the average length use information about geodesics that penetrate deeper into the bulk than those relevant for entanglement, so both of them could play a role in bulk reconstruction. For instance, while recent work \cite{Headrick:2014cta} has connected the RT formula to questions of causality like those studied in \cite{Gao:2000ga, Hernandez-Cuenca:2021eyf, Berenstein:2020cll}, the former cannot always provide complete information about the latter due to its reliance on minimal surfaces. It may therefore be natural to turn to spacetimes like conical $\mathrm{AdS}_3$, where multiple geodesics can go deeper into the bulk spacetime, to make such discussions more precise.



\acknowledgments

It is a pleasure to thank Gary Horowitz, Don Marolf, Xi Dong, and Joan Simon for useful discussions; Adolfo Holguin for advice on hypergeometric functions and comments on an early draft of the manuscript; and Liam Fitzpatrick, Henry Maxfield, and Sean McBride for helpful conversations about conformal field theory.  D.B. and D.G. were supported in part by the Department of Energy under grant DE-SC 0011702. Z.L. was supported by the Worster Summer 2021 Research Fellowship at UCSB.

\appendix

\section{Normalization of the Radial Modes}
\label{sec:normalization}

Here we evaluate the integral (\ref{eqn:mode-norm}), which gives the norm $\big\Vert \phi_{n\ell} \big\Vert^2$ of the field modes. 

We will take $\ell \geq 0$ for simplicity, but an identical argument goes through for $\ell < 0$. The integral is finite only when the quantization condition (\ref{eqn:quantization}) is satisfied. The radial modes are given by (\ref{eqn:radial-modes}), so the integral we seek to evaluate is
\begin{align}
\label{eqn:normalization1}
\big\Vert \phi_{n\ell} \big\Vert^2 =
4\pi\w \mathcal{N}_{n\ell}^2 \int_0^{\infty} \dd r\, r^{1 + 2N\ell} \p{r^2 + \frac{1}{N^2}}^{N\w - 1} \big| \mathcal{F}_{n\ell}(r)\big|^2.
\end{align}

The goal will be to massage the integral into a form that \textit{Mathematica} is capable of evaluating directly. We begin by changing variables to $x = Nr$ to scale $N$ out of the integral. We have $r^2 + \frac{1}{N^2} = \frac{1}{N^2}\p{x^2 + 1}$ and $\dd r = N^{-1} \dd x$, so the norm becomes
\begin{align}
\label{eqn:normalization2}
\big\Vert \phi_{n\ell}\big\Vert^2 = \underbrace{4\pi \w \mathcal{N}_{n\ell}^2 N^{-2N\p{\ell + \w}}}_{A_{n\ell}}
\int_0^{\infty} \dd x\, x^{1 + 2N\ell} \p{x^2 + 1}^{N\w - 1} \Big| \mathcal{F}_{n\ell}\big(\tfrac{x}{N} \big)\Big|^2.
\end{align}
Next, let $z = -x^2$, so that $x^2 + 1 = 1 - z$ and $\dd z = -2x\, \dd x$. Then we obtain
\begin{equation}
\label{eqn:normalization3}
\begin{aligned}
\big\Vert \phi_{n\ell}\big\Vert^2 &= \frac{A_{n\ell}}{2} \int_{-\infty}^0 \dd z\, \p{-z}^{N\ell} \p{1-z}^{N\w - 1}
\big| \mathcal{F}_{n\ell}(z) \big|^2, \\[6pt]
\mathcal{F}_{n\ell}(z) &\equiv {}_2 F_1 \big[1 + n + N\ell,\, \Delta + n + N\ell;\, 1 + N\ell;\, z \big].
\end{aligned}
\end{equation}
The following Kummer relation \cite{Gradshteyn:abc} proves useful:
\begin{equation}
\label{eqn:kummer}
\begin{aligned}
{}_2 F_1 \big[\a,\,\b;\,\g;\,z \big] &= \p{1-z}^{\g-\a-\b} {}_2 F_1 \big[\g-\a,\,\g-\b;\,\g;\,z \big] \implies \\[6pt]
\mathcal{F}_{n\ell}(z) &= \p{1-z}^{-N\w} {}_2 F_1 \big[-n,\, 1 - \Delta - n;\, 1 + N\ell;\, z \big].
\end{aligned}
\end{equation}
We apply (\ref{eqn:kummer}) to one of the two hypergeometric functions in the integrand of (\ref{eqn:normalization3}). Finally, we change variables $z \too -z$, whence the norm becomes
\begin{equation}
\label{eqn:normalization4}
\begin{aligned}
\big\Vert \phi_{n\ell} \big\Vert^2 = \frac{A_{n\ell}}{2} \int_0^{\infty} \dd z \p{\frac{z^{N\ell}}{1+z}} \bigg( &{}_2 F_1 \big[1 + n + N\ell,\, \Delta + n + N\ell,\, 1 + N\ell,\, -z \big] \times \Big. \\ \Big. \times\,
&{}_2 F_1 \big[-n,\, 1 - \Delta - n,\, 1+N\ell,\, -z \big] \bigg).
\end{aligned}
\end{equation}

Amazingly, \textit{Mathematica} can evaluate this integral explicitly. The result is initially expressed in terms of Gamma functions and regularized hypergeometric ${}_2 F_1$ functions:
\begin{equation}
\begin{aligned}
\label{eqn:integral-answer}
\big\Vert \phi_{n\ell} \big\Vert^2 &= \frac{A_{n\ell}}{2} \p{\frac{\pi}{\sin\p{\pi \Delta}}} \p{\frac{\G(N\ell + 1)^2\, \G(N\w)}{\G(N\ell + n + 1)^2 \G(N\w - n)^2}} B_{n\ell}, \\[6pt]
A_{n\ell} &= 4\pi\w N^{-2N\p{\ell + \w} - 1} \mathcal{N}_{n\ell}^2, \\[6pt]
B_{n\ell} &= \G(\Delta+n) \G(N\w - n) {}_2 \tilde{F}_1 \big[\Delta+n,\, n+N\ell+1,\, \Delta,\, 1\big] - \\ &- \G(n+1) \G(n+N\ell+1) {}_2 \tilde{F}_1 \big[n+1,\, n+N\ell+1,\, 2-\Delta,\, 1\big].
\end{aligned}
\end{equation}
Here the regularized hypergeometric function ${}_2 \tilde{F}_1$ is defined and evaluated at $z=1$ by
\begin{align}
\label{eqn:regularized}
{}_2 \tilde{F}_1 [\a,\,\b;\,\g;\,z] \equiv \frac{{}_2 F_1 [\a,\,\b;\,\g;\,z]}{\G(\g)}, \qquad
{}_2 \tilde{F}_1 \big[\a,\,\b;\,\g;\, 1] = \frac{\G(\g - \a - \b)}{\G(\g - \a) \G(\g - \b)}.
\end{align}
But the ${}_2 \tilde{F}_1$ function in the first term of $B_{n\ell}$ vanishes: the denominator of its expression via (\ref{eqn:regularized}) contains the singular factor $\G(-n)$. Thus $\big\Vert \phi_{n\ell} \big\Vert^2$ reduces to a product of Gamma functions that simplify, using $\G(z) \G(1-z) = \frac{\pi}{\sin\p{\pi z}}$, to the desired expression (\ref{eqn:normalization-constant}):
\begin{equation}
\begin{aligned}
\big\Vert \phi_{n\ell}\big\Vert^2 &= \frac{2\pi \G(n+1) \G(1 + N\ell)^2 \G(n + \Delta)}{\G(1 + n + N\ell) \G(\Delta + n + N\ell)} N^{-2\Delta - 4\p{n + N\ell} - 1} \mathcal{N}_{n\ell}^2 = 1 \implies \\[6pt] \mathcal{N}_{n\ell}^2 &= \frac{\G(1 + n + N\ell) \G(\Delta + n + N\ell)}{2\pi \G(n+1) \G(1 + N\ell)^2 \G(n + \Delta)} N^{2\Delta + 4\p{n + N\ell} + 1}.
\end{aligned}
\end{equation}

\section{The Length of Spacelike Geodesics}
\label{sec:geodesic-length}

Here we compute the proper length of spacelike geodesics in the conical $\mathrm{AdS}_3$ spacetime. In coordinates $x^{\mu} = (t,r,\th)$, with $t \in \R$, $r \in \R_+$, $\th \in [0,2\pi)$, and $N \in \R_+$, the metric is
\begin{equation}
\begin{aligned}
\label{eqn:metric-again}
\dd s^2 &= g_{\mu\nu} \dd x^{\mu} \dd x^{\nu} =
-\p{r^2 + \frac{1}{N^2}} \dd t^2 + \p{r^2 + \frac{1}{N^2}}^{-1} \dd r^2 + r^2\, \dd\th^2 = \\ &= - H(r)\, \dd t^2 + H(r)^{-1}\, \dd r^2 + r^2\, \dd \th^2, \qquad H(r) = r^2 + \frac{1}{N^2}.
\end{aligned}
\end{equation}
Spacelike geodesics are constrained by normalization requirement $g_{\mu\nu} \dot{x}^{\mu} \dot{x}^{\nu} = +1$, where a dot denotes differentiation with respect to the proper length $s$. In our case, we have
\begin{align}
\label{eqn:spacelike}
-H(r) \dot{t}^2 + H(r)^{-1} \dot{r}^2 + r^2 \dot{\th}^2 = 1.
\end{align}
The metric (\ref{eqn:metric-again}) possesses two Killing vector fields, $\xi^{\mu} = \p{\partial_t}^{\mu}$ and $\eta^{\mu} = \p{\partial_{\th}}^{\mu}$. These symmetries give rise to a conserved energy and angular momentum, respectively:
\begin{align}
\label{eqn:conserved}
E \equiv -g_{\mu\nu} \xi^{\mu} \dot{x}^{\nu} = H(r) \dot{t}, \qquad L \equiv g_{\mu\nu} \eta^{\mu} \dot{x}^{\nu} = r^2 \dot{\th}.
\end{align}
We substitute $E$ and $L$ into the spacelike condition (\ref{eqn:spacelike}) to obtain the geodesic equation for $r(s)$ in the form of a Newtonian problem for a particle in a one-dimensional effective potential $V_{\mathrm{eff}}(r)$. Thus the geodesic equation for $r$ is reduced to quadratures:
\begin{align}
\label{eqn:newtonian}
E^2 = \dot{r}^2 + H(r) \p{\frac{L^2}{r^2} - 1} \equiv \dot{r}^2 + V_{\mathrm{eff}}(r) \implies \dot{r} = \dv{r}{s} =  \sqrt{E^2 - V_{\mathrm{eff}}(r)}.
\end{align}
The two other geodesic equations, which follow from (\ref{eqn:conserved}), are also separable:
\begin{align}
\dot{t} = \dv{t}{s} = \frac{E}{H(r)}, \qquad
\dot{\th} = \dv{\th}{s} = \frac{L}{r^2}.
\end{align}

To find the proper length of a geodesic that starts and ends on the boundary, we integrate (\ref{eqn:newtonian}) from a large IR cutoff $r = r_{\infty}$ to the radial position $r = r_*$ of the deepest bulk point on the geodesic---the turning point---and then back from $r_*$ to $r_{\infty}$. The turning point $r_*$ is somewhat like an impact parameter in scattering theory, and is determined by the ratio of the geodesic's angular momentum to its energy. Formally, $r_*$ is defined by
\begin{align}
\label{eqn:turning-pt}
E^2 = V_{\mathrm{eff}}(r_*) = \p{r_*^2 + \frac{1}{N^2}}\p{\frac{L^2}{r_*^2} - 1}.
\end{align}
This equation can be solved explicitly for $r_*$, but it will be more useful to view (\ref{eqn:turning-pt}) as an equation for $E$ in terms of $r_*$. We substitute it into (\ref{eqn:newtonian}) and integrate to find the length:
\begin{equation}
\label{eqn:length-integral}
\begin{aligned}
\L &= \int_0^{\L} \dd s = 2\int_{r_*}^{r_{\infty}} \frac{\dd r}{\sqrt{E^2 - V_{\mathrm{eff}}(r)}} = 2\int_{r_*}^{r_{\infty}} \frac{\dd r}{\sqrt{V_{\mathrm{eff}}(r_*) - V_{\mathrm{eff}}(r)}} = \\ &= 2 \sinh^{-1} \Bigg(N r_* \sqrt{\frac{r_{\infty}^2 - r_*^2}{L^2 + \p{Nr_*^2}^2}}\,\Bigg) \approx 2 \ln\p{r_{\infty}} + 2 \ln\Bigg(\frac{2N r_*}{\sqrt{L^2 + \p{Nr_*^2}^2}} \Bigg).
\end{aligned}
\end{equation}
Here we have assumed that the cutoff $r_{\infty}$ is much larger than all other scales involved.

To parametrize the length by $t$ and $\th$, we will determine $r_* = r_*(t,\th)$ and substitute the result into (\ref{eqn:length-integral}). For this, we will need the following useful relations:
\begin{align}
\dv{r}{t} = \dv{r}{s} \dv{s}{t} = \frac{H(r)}{E} \sqrt{E^2 - V_{\mathrm{eff}}(r)}, \qquad
\dv{r}{\th} = \dv{r}{s} \dv{s}{\th} = \frac{r^2}{L} \sqrt{E^2 - V_{\mathrm{eff}}(r)}.
\end{align}
Both of these differential equations are separable. After substituting $E^2 = V_{\mathrm{eff}}(r_*)$, they can be integrated in closed form without the need for a radial cutoff:
\begin{subequations}
\begin{align}
\label{eqn:integral-t}
t &= \int_0^t \dd t' = 2\int_{r_*}^{\infty} \frac{\sqrt{V_{\mathrm{eff}}(r_*)}\, \dd r}{H(r) \sqrt{V_{\mathrm{eff}}(r_*) - V_{\mathrm{eff}}(r)}} = 2N \tan^{-1}\p{\sqrt{\frac{L^2/r_*^2 - 1}{1 + \p{Nr_*}^2}}}, \\
\label{eqn:integral-th}
\th &= \int_0^{\th} \dd \th' = 2 \int_{r_*}^{\infty} \frac{L\, \dd r}{r^2 \sqrt{V_{\mathrm{eff}}(r_*) - V_{\mathrm{eff}}(r)}} = 2N \tan^{-1} \p{\frac{L}{Nr_*^2}}.
\end{align}
\end{subequations}
Now we solve (\ref{eqn:integral-th}) for $L$, substitute it into (\ref{eqn:integral-t}), and invert to find $r_*$:
\begin{align}
\label{eqn:L-and-rst}
L = Nr_*^2 \tan\p{\frac{\th}{2N}} \implies 
r_*^2 = \frac{\frac{1}{N^2} \p{1 + \cos\p{\frac{\th}{N}}}}{\cos\p{\frac{t}{N}} + \cos\p{\frac{\th}{N}}}.
\end{align}
Finally, we can substitute (\ref{eqn:L-and-rst}) directly into (\ref{eqn:length-integral}). After some algebra, we find\footnote{The absence of the term $2\ln\p{r_{\infty}}$ in the main text indicates our use of the renormalized length.}
\begin{align}
\label{eqn:time-dep-length}
\L = \L[\g(t,\th)] = 2 \ln\p{r_{\infty}} + \ln \qty[2N^2 \p{\cos\p{\frac{t}{N}} - \cos\p{\frac{\th}{N}}}].
\end{align}
At constant time ($t=0$), the result above reduces to
\begin{align}
\L[\g(0,\th)] \equiv \L[\g(\th)] = 2 \ln\p{r_{\infty}} + 2 \ln\qty[2N \sin\p{\frac{\th}{2N}}].
\end{align}

Because we have worked locally, the result (\ref{eqn:time-dep-length}) is insensitive to global features like the presence of multiple bulk geodesics, and only gives the length of the minimal bulk geodesic for angles $\th < \pi$. For larger separations $\th \geq \pi$, one substitutes $\th \too 2\pi - \th$ in (\ref{eqn:time-dep-length}) to find the minimal geodesic length. Following the discussion in \S\ref{sec:geodesic-approximation}, one can find the proper length of all $N$ bulk geodesics $\mathrm{AdS}_3/\Z_N$ by passing to the $\mathrm{AdS}_3$ cover and considering the geodesics between the initial point and the $\Z_N$ images of the final point. Thus (\ref{eqn:time-dep-length}) still gives the length of the $k$th geodesic, but with $\th$ replaced by $\th + 2\pi k$.

\end{document}

%% file: cylinder_2.pdf_tex
\begingroup%
  \makeatletter%
  \providecommand\color[2][]{%
    \errmessage{(Inkscape) Color is used for the text in Inkscape, but the package 'color.sty' is not loaded}%
    \renewcommand\color[2][]{}%
  }%
  \providecommand\transparent[1]{%
    \errmessage{(Inkscape) Transparency is used (non-zero) for the text in Inkscape, but the package 'transparent.sty' is not loaded}%
    \renewcommand\transparent[1]{}%
  }%
  \providecommand\rotatebox[2]{#2}%
  \newcommand*\fsize{\dimexpr\f@size pt\relax}%
  \newcommand*\lineheight[1]{\fontsize{\fsize}{#1\fsize}\selectfont}%
  \ifx\svgwidth\undefined%
    \setlength{\unitlength}{524.33863986bp}%
    \ifx\svgscale\undefined%
      \relax%
    \else%
      \setlength{\unitlength}{\unitlength * \real{\svgscale}}%
    \fi%
  \else%
    \setlength{\unitlength}{\svgwidth}%
  \fi%
  \global\let\svgwidth\undefined%
  \global\let\svgscale\undefined%
  \makeatother%
  \begin{picture}(1,0.47559437)%
    \lineheight{1}%
    \setlength\tabcolsep{0pt}%
    \put(0,0){\includegraphics[width=\unitlength,page=1]{cylinder_2.pdf}}%
    \put(0.35488983,0.2500146){\color[rgb]{0,0,0}\makebox(0,0)[lt]{\lineheight{1.25}\smash{\begin{tabular}[t]{l}$\scalebox{1.5}{$\color{NavyBlue}{\phi(0,0)}$}$\end{tabular}}}}%
    \put(0.24493091,0.17058653){\color[rgb]{0,0,0}\makebox(0,0)[lt]{\lineheight{1.25}\smash{\begin{tabular}[t]{l}$\scalebox{1.5}{$\color{NavyBlue}{\phi(0, \th)}$}$\end{tabular}}}}%
    \put(0.08018626,0.0698153){\color[rgb]{0,0,0}\makebox(0,0)[lt]{\lineheight{1.25}\smash{\begin{tabular}[t]{l}$\scalebox{1.5}{$\color{Maroon}{\mathcal{O}_H(t = -\infty)}$}$\end{tabular}}}}%
    \put(0.08135018,0.43111966){\color[rgb]{0,0,0}\makebox(0,0)[lt]{\lineheight{1.25}\smash{\begin{tabular}[t]{l}$\scalebox{1.5}{$\color{Maroon}{\mathcal{O}_H(t = +\infty)}$}$\end{tabular}}}}%
    \put(0.73315186,0.21872231){\color[rgb]{0,0,0}\makebox(0,0)[lt]{\lineheight{1.25}\smash{\begin{tabular}[t]{l}$\scalebox{1.5}{$\color{Maroon}{\mathcal{O}_H(0)}$}$\end{tabular}}}}%
    \put(0.82098231,0.37388514){\color[rgb]{0,0,0}\makebox(0,0)[lt]{\lineheight{1.25}\smash{\begin{tabular}[t]{l}$\scalebox{1.5}{$\color{Maroon}{\mathcal{O}_H(\infty)}$}$\end{tabular}}}}%
    \put(0.87899299,0.24653987){\color[rgb]{0,0,0}\makebox(0,0)[lt]{\lineheight{1.25}\smash{\begin{tabular}[t]{l}$\scalebox{1.5}{$\color{NavyBlue}{\mathcal{O}_L(1)}$}$\end{tabular}}}}%
    \put(0.81153714,0.14741785){\color[rgb]{0,0,0}\makebox(0,0)[lt]{\lineheight{1.25}\smash{\begin{tabular}[t]{l}$\scalebox{1.5}{$\color{NavyBlue}{\mathcal{O}_L(z, \bar{z})}$}$\end{tabular}}}}%
    \put(0.38647819,0.35115566){\color[rgb]{0,0,0}\makebox(0,0)[lt]{\lineheight{1.25}\smash{\begin{tabular}[t]{l}$\scalebox{1.5}{$z = e^{\t + i\th}$}$\end{tabular}}}}%
    \put(0.35656944,0.32164901){\color[rgb]{0,0,0}\makebox(0,0)[lt]{\lineheight{1.25}\smash{\begin{tabular}[t]{l}$\scalebox{2.5}{$\xrightarrow{\hspace*{1.2cm}}$}$\end{tabular}}}}%
  \end{picture}%
\endgroup%

%% file: ConicalAdS_final.bbl
\begin{thebibliography}{99}

\bibitem{Maldacena:1997re}
J.~M.~Maldacena,
``The Large N limit of superconformal field theories and supergravity,''
Adv. Theor. Math. Phys. \textbf{2}, 231-252 (1998)
doi:10.1023/A:1026654312961
[arXiv:hep-th/9711200 [hep-th]].

\bibitem{Witten:1998qj}
E.~Witten,
``Anti-de Sitter space and holography,''
Adv. Theor. Math. Phys. \textbf{2}, 253-291 (1998)
doi:10.4310/ATMP.1998.v2.n2.a2
[arXiv:hep-th/9802150 [hep-th]].

\bibitem{Gubser:1998bc}
S.~S.~Gubser, I.~R.~Klebanov and A.~M.~Polyakov,
``Gauge theory correlators from noncritical string theory,''
Phys. Lett. B \textbf{428}, 105-114 (1998)
doi:10.1016/S0370-2693(98)00377-3
[arXiv:hep-th/9802109 [hep-th]].

\bibitem{Hawking:1982dh}
S.~W.~Hawking and D.~N.~Page,
``Thermodynamics of Black Holes in anti-De Sitter Space,''
Commun. Math. Phys. \textbf{87}, 577 (1983)
doi:10.1007/BF01208266

\bibitem{Witten:1998zw}
E.~Witten,
``Anti-de Sitter space, thermal phase transition, and confinement in gauge theories,''
Adv. Theor. Math. Phys. \textbf{2}, 505-532 (1998)
doi:10.4310/ATMP.1998.v2.n3.a3
[arXiv:hep-th/9803131 [hep-th]].

\bibitem{Deser:1983nh}
S.~Deser and R.~Jackiw,
``Three-Dimensional Cosmological Gravity: Dynamics of Constant Curvature,''
Annals Phys. \textbf{153}, 405-416 (1984)
doi:10.1016/0003-4916(84)90025-3

\bibitem{Banados:1992wn}
M.~Banados, C.~Teitelboim and J.~Zanelli,
``The Black hole in three-dimensional space-time,''
Phys. Rev. Lett. \textbf{69}, 1849-1851 (1992)
doi:10.1103/PhysRevLett.69.1849
[arXiv:hep-th/9204099 [hep-th]].

\bibitem{Avis:1977yn}
S.~J.~Avis, C.~J.~Isham and D.~Storey,
``Quantum Field Theory in anti-De Sitter Space-Time,''
Phys. Rev. D \textbf{18}, 3565 (1978)
doi:10.1103/PhysRevD.18.3565

\bibitem{Balasubramanian:1998sn}
V.~Balasubramanian, P.~Kraus and A.~E.~Lawrence,
``Bulk versus boundary dynamics in anti-de Sitter space-time,''
Phys. Rev. D \textbf{59}, 046003 (1999)
doi:10.1103/PhysRevD.59.046003
[arXiv:hep-th/9805171 [hep-th]].

\bibitem{Freedman:1998tz}
D.~Z.~Freedman, S.~D.~Mathur, A.~Matusis and L.~Rastelli,
``Correlation functions in the CFT(d) / AdS(d+1) correspondence,''
Nucl. Phys. B \textbf{546}, 96-118 (1999)
doi:10.1016/S0550-3213(99)00053-X
[arXiv:hep-th/9804058 [hep-th]].

\bibitem{Berenstein:1998ij}
D.~E.~Berenstein, R.~Corrado, W.~Fischler and J.~M.~Maldacena,
``The Operator product expansion for Wilson loops and surfaces in the large N limit,''
Phys. Rev. D \textbf{59}, 105023 (1999)
doi:10.1103/PhysRevD.59.105023
[arXiv:hep-th/9809188 [hep-th]].

\bibitem{Arefeva:2014aoe}
I.~Y.~Arefeva and A.~A.~Bagrov,
``Holographic dual of a conical defect,''
Teor. Mat. Fiz. \textbf{182}, no.1, 3-27 (2014)
https://doi.org/10.4213/tmf8747.

\bibitem{Banks:1998dd}
T.~Banks, M.~R.~Douglas, G.~T.~Horowitz and E.~J.~Martinec,
``AdS dynamics from conformal field theory,''
[arXiv:hep-th/9808016 [hep-th]].

\bibitem{Fitzpatrick:2014vua}
A.~L.~Fitzpatrick, J.~Kaplan and M.~T.~Walters,
``Universality of Long-Distance AdS Physics from the CFT Bootstrap,''
JHEP \textbf{08}, 145 (2014)
doi:10.1007/JHEP08(2014)145
[arXiv:1403.6829 [hep-th]].

\bibitem{Fitzpatrick:2015zha}
A.~L.~Fitzpatrick, J.~Kaplan and M.~T.~Walters,
``Virasoro Conformal Blocks and Thermality from Classical Background Fields,''
JHEP \textbf{11}, 200 (2015)
doi:10.1007/JHEP11(2015)200
[arXiv:1501.05315 [hep-th]].

\bibitem{Calabrese:2009qy}
P.~Calabrese and J.~Cardy,
``Entanglement entropy and conformal field theory,''
J. Phys. A \textbf{42}, 504005 (2009)
doi:10.1088/1751-8113/42/50/504005
[arXiv:0905.4013 [cond-mat.stat-mech]].

\bibitem{Lewkowycz:2013nqa}
A.~Lewkowycz and J.~Maldacena,
``Generalized gravitational entropy,''
JHEP \textbf{08}, 090 (2013)
doi:10.1007/JHEP08(2013)090
[arXiv:1304.4926 [hep-th]].

\bibitem{Benjamin:2020mfz}
N.~Benjamin, S.~Collier and A.~Maloney,
``Pure Gravity and Conical Defects,''
JHEP \textbf{09}, 034 (2020)
doi:10.1007/JHEP09(2020)034
[arXiv:2004.14428 [hep-th]].

\bibitem{Martinec:2001cf}
E.~J.~Martinec and W.~McElgin,
``String theory on AdS orbifolds,''
JHEP \textbf{04}, 029 (2002)
doi:10.1088/1126-6708/2002/04/029
[arXiv:hep-th/0106171 [hep-th]].

\bibitem{Balasubramanian:2005qu}
V.~Balasubramanian, P.~Kraus and M.~Shigemori,
``Massless black holes and black rings as effective geometries of the D1-D5 system,''
Class. Quant. Grav. \textbf{22}, 4803-4838 (2005)
doi:10.1088/0264-9381/22/22/010
[arXiv:hep-th/0508110 [hep-th]].

\bibitem{Lunin:2001jy}
O.~Lunin and S.~D.~Mathur,
``AdS / CFT duality and the black hole information paradox,''
Nucl. Phys. B \textbf{623}, 342-394 (2002)
doi:10.1016/S0550-3213(01)00620-4
[arXiv:hep-th/0109154 [hep-th]].

\bibitem{Balasubramanian:2014sra}
V.~Balasubramanian, B.~D.~Chowdhury, B.~Czech and J.~de Boer,
``Entwinement and the emergence of spacetime,''
JHEP \textbf{01}, 048 (2015)
doi:10.1007/JHEP01(2015)048
[arXiv:1406.5859 [hep-th]].

\bibitem{Giusto:2004ip}
S.~Giusto, S.~D.~Mathur and A.~Saxena,
``3-charge geometries and their CFT duals,''
Nucl. Phys. B \textbf{710}, 425-463 (2005)
doi:10.1016/j.nuclphysb.2005.01.009
[arXiv:hep-th/0406103 [hep-th]].

\bibitem{Giusto:2012yz}
S.~Giusto, O.~Lunin, S.~D.~Mathur and D.~Turton,
``D1-D5-P microstates at the cap,''
JHEP \textbf{02}, 050 (2013)
doi:10.1007/JHEP02(2013)050
[arXiv:1211.0306 [hep-th]].

\bibitem{Galliani:2016cai}
A.~Galliani, S.~Giusto, E.~Moscato and R.~Russo,
``Correlators at large c without information loss,''
JHEP \textbf{09}, 065 (2016)
doi:10.1007/JHEP09(2016)065
[arXiv:1606.01119 [hep-th]].

\bibitem{Giusto:2020mup}
S.~Giusto, M.~R.~R.~Hughes and R.~Russo,
``The Regge limit of AdS$_{3}$ holographic correlators,''
JHEP \textbf{11}, 018 (2020)
doi:10.1007/JHEP11(2020)018
[arXiv:2007.12118 [hep-th]].

\bibitem{Kulaxizi:2018dxo}
M.~Kulaxizi, G.~S.~Ng and A.~Parnachev,
SciPost Phys. \textbf{6}, no.6, 065 (2019)
doi:10.21468/SciPostPhys.6.6.065
[arXiv:1812.03120 [hep-th]].

\bibitem{Karlsson:2019qfi}
R.~Karlsson, M.~Kulaxizi, A.~Parnachev and P.~Tadi\'c,
JHEP \textbf{10}, 046 (2019)
doi:10.1007/JHEP10(2019)046
[arXiv:1904.00060 [hep-th]].

\bibitem{Ageev:2015qbz}
D.~S.~Ageev, I.~Y.~Aref'eva and M.~D.~Tikhanovskaya,
``(1+1)-Correlators and moving massive defects,''
Theor. Math. Phys. \textbf{188}, no.1, 1038-1068 (2016)
doi:10.1134/S0040577916070060
[arXiv:1512.03362 [hep-th]].

\bibitem{Arefeva:2016nic}
I.~Y.~Aref'eva, M.~A.~Khramtsov and M.~D.~Tikhanovskaya,
``Improved image method for a holographic description of conical defects,''
Theor. Math. Phys. \textbf{189}, no.2, 1660-1672 (2016)
doi:10.1134/S0040577916110106
[arXiv:1604.08905 [hep-th]].

\bibitem{Arefeva:2016wek}
I.~Y.~Arefeva and M.~A.~Khramtsov,
``AdS/CFT prescription for angle-deficit space and winding geodesics,''
JHEP \textbf{04}, 121 (2016)
doi:10.1007/JHEP04(2016)121
[arXiv:1601.02008 [hep-th]].

\bibitem{Collier:2018exn}
S.~Collier, Y.~Gobeil, H.~Maxfield and E.~Perlmutter,
``Quantum Regge Trajectories and the Virasoro Analytic Bootstrap,''
JHEP \textbf{05}, 212 (2019)
doi:10.1007/JHEP05(2019)212
[arXiv:1811.05710 [hep-th]].

\bibitem{Klebanov:1999tb}
I.~R.~Klebanov and E.~Witten,
``AdS / CFT correspondence and symmetry breaking,''
Nucl. Phys. B \textbf{556}, 89-114 (1999)
doi:10.1016/S0550-3213(99)00387-9
[arXiv:hep-th/9905104 [hep-th]].

\bibitem{Breitenlohner:1982jf}
P.~Breitenlohner and D.~Z.~Freedman,
``Stability in Gauged Extended Supergravity,''
Annals Phys. \textbf{144}, 249 (1982)
doi:10.1016/0003-4916(82)90116-6.

\bibitem{Friedlander:abc}
F.~G.~Friedlander, \textit{The Wave Equation on a Curved Spacetime}, Cambridge University Press, New York (1975). ISBN: 978-0521136365.

\bibitem{F.TRICOMI}
F.~Tricomi and A.~Eed\'elyi,
``The Asymptotic Expansion of a Ratio of Gamma Functions,''
Pacific Journal of Mathematics, Vol. 1, No. 1 (1951).

\bibitem{Balasubramanian:1999zv}
V.~Balasubramanian and S.~F.~Ross,
``Holographic particle detection,''
Phys. Rev. D \textbf{61}, 044007 (2000)
doi:10.1103/PhysRevD.61.044007
[arXiv:hep-th/9906226 [hep-th]].

\bibitem{Ballmann}
W.~Ballmann, M.~Brin and K.~Burns,
``On surfaces with no conjugate points,''
Journal of Differential Geometry. March (1987)
doi:10.4310/jdg/1214440852

\bibitem{Berenstein:2002ke}
D.~Berenstein, C.~P.~Herzog and I.~R.~Klebanov,
``Baryon spectra and AdS /CFT correspondence,''
JHEP \textbf{06}, 047 (2002)
doi:10.1088/1126-6708/2002/06/047
[arXiv:hep-th/0202150 [hep-th]].

\bibitem{Calabrese:2010he}
P.~Calabrese, J.~Cardy and E.~Tonni,
``Entanglement entropy of two disjoint intervals in conformal field theory II,''
J. Stat. Mech. \textbf{1101}, P01021 (2011)
doi:10.1088/1742-5468/2011/01/P01021
[arXiv:1011.5482 [hep-th]].

\bibitem{Hardy:abc}
G.~H.~Hardy, ``On two theorems of F. Carlson and S. Wigert,''
Acta Mathematica \textbf{42}: 327--329 (1920) doi:10.1007/bf02404414.

\bibitem{Brown:1986nw}
J.~D.~Brown and M.~Henneaux,
``Central Charges in the Canonical Realization of Asymptotic Symmetries: An Example from Three-Dimensional Gravity,''
Commun. Math. Phys. \textbf{104}, 207-226 (1986)
doi:10.1007/BF01211590

\bibitem{Berenstein:2021pxv}
D.~Berenstein and R.~B.~de Zoysa,
``Operator product expansions and recoil,''
Phys. Rev. D \textbf{105}, no.2, 026019 (2022)
doi:10.1103/PhysRevD.105.026019
[arXiv:2110.15297 [hep-th]].

\bibitem{Berenstein:2014cia}
D.~Berenstein and A.~Miller,
``Conformal perturbation theory, dimensional regularization, and AdS/CFT correspondence,''
Phys. Rev. D \textbf{90}, no.8, 086011 (2014)
doi:10.1103/PhysRevD.90.086011
[arXiv:1406.4142 [hep-th]].

\bibitem{Berenstein:2019tcs}
D.~Berenstein and J.~Sim\'on,
``Localized states in global AdS space,''
Phys. Rev. D \textbf{101}, no.4, 046026 (2020)
doi:10.1103/PhysRevD.101.046026
[arXiv:1910.10227 [hep-th]].

\bibitem{Dolan:2000ut}
F.~A.~Dolan and H.~Osborn,
``Conformal four point functions and the operator product expansion,''
Nucl. Phys. B \textbf{599}, 459-496 (2001)
doi:10.1016/S0550-3213(01)00013-X
[arXiv:hep-th/0011040 [hep-th]].

\bibitem{Watson:1918abc}
G.~N.~Watson, ``Asymptotic Expansions of Hypergeometric Functions,'' Trans. Cambridge Philos. Soc., \textbf{22}, 277-308 (1918).

\bibitem{Zamolodchikov:1986gt}
A.~B.~Zamolodchikov,
``Irreversibility of the Flux of the Renormalization Group in a 2D Field Theory,''
JETP Lett. \textbf{43}, 730-732 (1986)

\bibitem{Hijano:2015qja}
E.~Hijano, P.~Kraus, E.~Perlmutter and R.~Snively,
``Semiclassical Virasoro blocks from AdS$_{3}$ gravity,''
JHEP \textbf{12}, 077 (2015)
doi:10.1007/JHEP12(2015)077
[arXiv:1508.04987 [hep-th]].
doi:10.1143/JPSJ.12.570.

\bibitem{Asplund:2014coa}
C.~T.~Asplund, A.~Bernamonti, F.~Galli and T.~Hartman,
``Holographic Entanglement Entropy from 2d CFT: Heavy States and Local Quenches,''
JHEP \textbf{02}, 171 (2015)
doi:10.1007/JHEP02(2015)171
[arXiv:1410.1392 [hep-th]].

\bibitem{Kubo:abc}
Kubo, Ryogo, 
``Statistical-Mechanical Theory of Irreversible Processes. I. General Theory and Simple Applications to Magnetic and Conduction Problems,'' 
J. Phys. Soc. Jpn. \textbf{12}: 570--586 (1957)

\bibitem{Berenstein:2016avf}
D.~Berenstein and A.~Miller,
``Logarithmic enhancements in conformal perturbation theory and their real time interpretation,''
Int. J. Mod. Phys. A \textbf{35}, no.29, 2050184 (2020)
doi:10.1142/S0217751X20501845
[arXiv:1607.01922 [hep-th]].

\bibitem{Berenstein:2020cll}
D.~Berenstein and D.~Grabovsky,
``The Tortoise and the Hare: A Causality Puzzle in AdS/CFT,''
Class. Quant. Grav. \textbf{38}, no.10, 105008 (2021)
doi:10.1088/1361-6382/abf1c7
[arXiv:2011.08934 [hep-th]].

\bibitem{Banerjee:2016qca}
P.~Banerjee, S.~Datta and R.~Sinha,
``Higher-point conformal blocks and entanglement entropy in heavy states,''
JHEP \textbf{05}, 127 (2016)
doi:10.1007/JHEP05(2016)127
[arXiv:1601.06794 [hep-th]].

\bibitem{Duetsch:2002hc}
M.~Duetsch and K.~H.~Rehren,
``Generalized free fields and the AdS - CFT correspondence,''
Annales Henri Poincare \textbf{4}, 613-635 (2003)
doi:10.1007/s00023-003-0141-9
[arXiv:math-ph/0209035 [math-ph]].

\bibitem{Ryu:2006ef}
S.~Ryu and T.~Takayanagi,
``Aspects of Holographic Entanglement Entropy,''
JHEP \textbf{08}, 045 (2006)
doi:10.1088/1126-6708/2006/08/045
[arXiv:hep-th/0605073 [hep-th]].

\bibitem{Fischetti:2014zja}
S.~Fischetti and D.~Marolf,
``Complex Entangling Surfaces for AdS and Lifshitz Black Holes?,''
Class. Quant. Grav. \textbf{31}, no.21, 214005 (2014)
doi:10.1088/0264-9381/31/21/214005
[arXiv:1407.2900 [hep-th]].

\bibitem{Marolf:2020vsi}
D.~Marolf, S.~Wang and Z.~Wang,
``Probing phase transitions of holographic entanglement entropy with fixed area states,''
JHEP \textbf{12}, 084 (2020)
doi:10.1007/JHEP12(2020)084
[arXiv:2006.10089 [hep-th]].

\bibitem{Dong:2020iod}
X.~Dong and H.~Wang,
``Enhanced corrections near holographic entanglement transitions: a chaotic case study,''
JHEP \textbf{11}, 007 (2020)
doi:10.1007/JHEP11(2020)007
[arXiv:2006.10051 [hep-th]].

\bibitem{Dong:2021oad}
X.~Dong, S.~McBride and W.~W.~Weng,
``Replica Wormholes and Holographic Entanglement Negativity,''
[arXiv:2110.11947 [hep-th]].

\bibitem{Headrick:abc}
M.~Headrick, ``Entanglement Entropy,'' Talk at KITP Conference: Quantum Fields Beyond Perturbation Theory (2014).

\bibitem{Berenstein:2022nlj}
D.~Berenstein and R.~Mancilla,
``Aspects of thermal one-point functions and response functions in AdS Black holes,''
[arXiv:2211.05144 [hep-th]].

\bibitem{Headrick:2014cta}
M.~Headrick, V.~E.~Hubeny, A.~Lawrence and M.~Rangamani,
``Causality \& holographic entanglement entropy,''
JHEP \textbf{12}, 162 (2014)
doi:10.1007/JHEP12(2014)162
[arXiv:1408.6300 [hep-th]].

\bibitem{Gao:2000ga}
S.~Gao and R.~M.~Wald,
``Theorems on gravitational time delay and related issues,''
Class. Quant. Grav. \textbf{17}, 4999-5008 (2000)
doi:10.1088/0264-9381/17/24/305
[arXiv:gr-qc/0007021 [gr-qc]].

\bibitem{Hernandez-Cuenca:2021eyf}
S.~Hern\'andez-Cuenca, G.~T.~Horowitz, G.~Trevi\~no and D.~Wang,
``Boundary Causality Violating Metrics in Holography,''
Phys. Rev. Lett. \textbf{127}, no.8, 8 (2021)
doi:10.1103/PhysRevLett.127.081603
[arXiv:2103.05014 [hep-th]].

\bibitem{Gradshteyn:abc}
I.~S.~Gradshteyn and I.~M.~Ryzhik, 
\textit{Table of Integrals, Series, and Products},
Edited by D. Zwillinger and V. Moll, Academic Press, New York, 8th edition (2015).

\end{thebibliography}
